\documentclass[12pt,a4paper,dvips]{article}
\usepackage{a4p}
\usepackage{cite,mcite}
\usepackage{graphicx,epsfig}
\usepackage{physics}
\usepackage{l3_title,ifthen}
\journalname{Phys. Lett. B}
\date{July 24, 2000}
\preprint{2000-104}
\newlength{\capindent}
\newlength{\capwidth}
\newlength{\figwidth}
\newcommand{\icaption}[2][!*!,!]{\hspace*{\capindent}%
  \begin{minipage}{\capwidth}
    \ifthenelse{\equal{#1}{!*!,!}}%
      {\caption{#2}}%
      {\caption[#1]{#2}}
  \end{minipage}}
\renewcommand{\eV}{\mathrm{e\kern -0.1em V}}
\renewcommand{\MeV}{\mathrm{Me\kern -0.1em V}}
\renewcommand{\GeV}{\mathrm{Ge\kern -0.1em V}}

\newcommand{\pz}{\ensuremath{\phantom{0}}}
\newcommand{\pzz}{\ensuremath{\phantom{00}}}
\newcommand{\GW}{\ensuremath{\Gamma_{\mathrm{W}}}}
\newcommand{\PW}{\ensuremath{\mathrm{W}}}
\newcommand{\WW}{\ensuremath{\mathrm{WW}}}
\newcommand{\SWW}{\ensuremath{\sigma_{\mathrm{WW}}}}
\newcommand{\WWG}{\ensuremath{\mathrm{W^+W^-}(\gamma)}}

\newcommand{\EE}{\ensuremath{\mathrm{e}^+\mathrm{e}^-}}
\newcommand{\MM}{\ensuremath{\mu^+\mu^-}}
\newcommand{\TT}{\ensuremath{\tau^+\tau^-}}

\newcommand{\QQ}{\ensuremath{\mathrm{q}\bar \mathrm{q}}}
\newcommand{\FF}{\ensuremath{\mathrm{f}\kern 0.05em \bar \mathrm{f}}}
\newcommand{\EN}{\ensuremath{\mathrm{e}\nu}}
\newcommand{\MN}{\ensuremath{\mu\nu}}
\newcommand{\TN}{\ensuremath{\tau\nu}}
\newcommand{\LN}{\ensuremath{\ell\nu}}
\newcommand{\WQQ}{\ensuremath{\PW\rightarrow \mathrm{qq}}}
\newcommand{\WFF}{\ensuremath{\PW\rightarrow \mathrm{f\kern 0.05em f}}}
\newcommand{\WEN}{\ensuremath{\PW\rightarrow \EN}}
\newcommand{\WMN}{\ensuremath{\PW\rightarrow \MN}}
\newcommand{\WTN}{\ensuremath{\PW\rightarrow \TN}}
\newcommand{\WLN}{\ensuremath{\PW\rightarrow \LN}}

\newcommand{\EEG}{\EE\ensuremath{(\gamma)}}
\newcommand{\EEGG}{\EE\ensuremath{\gamma (\gamma)}}
\newcommand{\MMG}{\MM\ensuremath{(\gamma)}}
\newcommand{\TTG}{\TT\ensuremath{(\gamma)}}

\newcommand{\QQG}{\QQ\ensuremath{(\gamma)}}

\newcommand{\EEEEG}{\EE\ensuremath{\rightarrow}\EEG}
\newcommand{\EEEEGG}{\EE\ensuremath{\rightarrow}\EEGG}
\newcommand{\EEMMG}{\EE\ensuremath{\rightarrow}\MMG}

\newcommand{\EEQQG}{\EE\ensuremath{\rightarrow}\QQG}

\newcommand{\QQEN}{\ensuremath{\mathrm{qq}\mathrm{e\nu}}}
\newcommand{\QQMN}{\ensuremath{\mathrm{qq}\mu\nu}}
\newcommand{\QQTN}{\ensuremath{\mathrm{qq}\tau\nu}}
\newcommand{\QQLN}{\ensuremath{\mathrm{qq}\ell\nu}}
\newcommand{\LNLN}{\ensuremath{\ell\nu\ell\nu}}
\newcommand{\ENEN}{\ensuremath{\mathrm{e}\nu\mathrm{e}\nu}}
\newcommand{\ENMN}{\ensuremath{\mathrm{e}\nu\mu\nu}}
\newcommand{\ENTN}{\ensuremath{\mathrm{e}\nu\tau\nu}}
\newcommand{\MNMN}{\ensuremath{\mu\nu\mu\nu}}
\newcommand{\MNTN}{\ensuremath{\mu\nu\tau\nu}}
\newcommand{\TNTN}{\ensuremath{\tau\nu\tau\nu}}
\newcommand{\QQQQ}{\ensuremath{\mathrm{qqqq}}}
\newcommand{\FFFF}{\ensuremath{\mathrm{f \kern 0.05em f \kern 0.05em f \kern 0.05em f}}}

\newcommand{\QQENG}{\QQEN\ensuremath{(\gamma)}}
\newcommand{\QQMNG}{\QQMN\ensuremath{(\gamma)}}
\newcommand{\QQTNG}{\QQTN\ensuremath{(\gamma)}}
\newcommand{\QQLNG}{\QQLN\ensuremath{(\gamma)}}
\newcommand{\LNLNG}{\LNLN\ensuremath{(\gamma)}}
\newcommand{\ENENG}{\ENEN\ensuremath{(\gamma)}}
\newcommand{\ENMNG}{\ENMN\ensuremath{(\gamma)}}
\newcommand{\ENTNG}{\ENTN\ensuremath{(\gamma)}}
\newcommand{\MNMNG}{\MNMN\ensuremath{(\gamma)}}
\newcommand{\MNTNG}{\MNTN\ensuremath{(\gamma)}}
\newcommand{\TNTNG}{\TNTN\ensuremath{(\gamma)}}
\newcommand{\QQQQG}{\QQQQ\ensuremath{(\gamma)}}
\newcommand{\FFFFG}{\FFFF\ensuremath{(\gamma)}}

\newcommand{\EEQQENG}{\EE\ensuremath{\rightarrow}\QQENG}
\newcommand{\EEQQMNG}{\EE\ensuremath{\rightarrow}\QQMNG}
\newcommand{\EEQQTNG}{\EE\ensuremath{\rightarrow}\QQTNG}
\newcommand{\EEQQLNG}{\EE\ensuremath{\rightarrow}\QQLNG}
\newcommand{\EELNLNG}{\EE\ensuremath{\rightarrow}\LNLNG}
\newcommand{\EEENENG}{\EE\ensuremath{\rightarrow}\ENENG}
\newcommand{\EEENMNG}{\EE\ensuremath{\rightarrow}\ENMNG}
\newcommand{\EEENTNG}{\EE\ensuremath{\rightarrow}\ENTNG}
\newcommand{\EEMNMNG}{\EE\ensuremath{\rightarrow}\MNMNG}
\newcommand{\EEMNTNG}{\EE\ensuremath{\rightarrow}\MNTNG}
\newcommand{\EETNTNG}{\EE\ensuremath{\rightarrow}\TNTNG}
\newcommand{\EEQQQQG}{\EE\ensuremath{\rightarrow}\QQQQG}
\newcommand{\EEFFFFG}{\EE\ensuremath{\rightarrow}\FFFFG}

\newcommand{\EEWWG}{\EE\ensuremath{\rightarrow}\WWG}
\newcommand{\EEWWFFFFG}{\EE\ensuremath{\rightarrow}\WW\ensuremath{\rightarrow}\FFFFG}

\def\p1mu{\mathrm{p^{\mu}_{1}}}

\begin{document}
\begin{titlepage}
\title{
       Measurement of the W-Pair Production Cross Section \\
       and W-Decay Branching Fractions      \\
       in \boldmath$\mathrm{e^+e^-}$ Interactions 
       at \boldmath$\sqrt{s}=189~\GeV$      \\}

\author{The L3 Collaboration}

%
%
\begin{abstract}
  The data collected by the L3 experiment at LEP at a centre-of-mass
  energy of $188.6~\GeV$ are used to measure the W-pair
  production cross section and the W-boson decay branching
  fractions. These data correspond to an integrated luminosity of 
  176.8~pb$^{-1}$.
  The total cross section for
  W-pair production, combining all final states, is measured to be $\SWW =
  16.24 \pm 0.37~(stat.) \pm 0.22~(syst.)$~pb.  Including our data
  collected at lower centre-of-mass energies, the hadronic branching
  fraction of the W-boson is determined to be $ B(\WQQ)= \left[ 68.20 \pm
  0.68~(stat.) \pm 0.33~(syst.)\right]~\%$.  The results agree with the
  Standard Model predictions.

\end{abstract}
%
%
\submitted
\end{titlepage}
%
%
\section{Introduction}

Since 1996 the electron-positron collider LEP at CERN has exceeded the
centre-of-mass energy required to produce W bosons in pairs, $\EE
\rightarrow \mathrm{W^+W^-}$.  To lowest order within the Standard
Model~\cite{standard_model,SM-2} (SM), three Feynman diagrams contribute to
W-pair production: $s$-channel $\gamma$ and Z-boson exchange and
$t$-channel $\nu_{\e}$ exchange, referred to as
CC03~\cite{CCNC,LEP2YRWW,LEP2YREG}.  The W boson decays into a
quark-antiquark pair, for example $\mathrm{W^-\rightarrow\bar u
  d~or~\bar c s}$, or a lepton-antilepton pair,
$\mathrm{W^-\rightarrow\ell^-\bar\nu_\ell}$ ($\ell=\e,\mu,\tau$), in
the following denoted as $\mathrm{qq}$, $\ell\nu$ or 
$\mathrm{f\kern 0.05em f}$ in general for both
W$^+$ and W$^-$ decays.

In 1998 the L3 detector~\cite{l3-01-new} collected an integrated luminosity of
176.8~pb$^{-1}$~\cite{l3-lumi96,l3-196} at a centre-of-mass energy, $\sqrt{s}$, of
$188.6~\GeV$, increasing our total W-pair statistics by a factor of
four.  Cross sections are measured for all four-fermion final states
arising in W-pair production. The total W-pair production
cross section is determined in a combined analysis.
For the W-decay branching fractions, the results are combined with our
previous measurements~\cite{L-XSEC-161-183}.    

\section{Analysis of  W-Pair Production}
\label{sec:selec}

The selections of W-pair events are based on the four-fermion final states
from  W-pair decays: $\LNLNG$, $\QQENG$,
$\QQMNG$, $\QQTNG$ and $\QQQQG$.  Additional contributions to
the production of these final
states from other neutral-current (NC) or charged-current (CC)
Feynman diagrams are small.  At the current level of statistical
accuracy the interference effects need to be taken into account only
for $\EELNLNG$ (CC56+NC56) and $\EEQQENG$
(CC20)~\cite{CCNC,LEP2YRWW,LEP2YREG}.

The selection criteria are
similar to those used for the data collected at $\sqrt{s}=183~\GeV$
in 1997~\cite{L-XSEC-161-183} but are 
adapted to the higher centre-of-mass energy.  
In the following, only the main
ideas and important changes are described.  Particular emphasis is
placed on the evaluation of systematic uncertainties.

For the identification of electrons, an energy deposition in 
the electro-magnetic calorimeter is required, matched in
azimuth with a track reconstructed in the central tracking system.  
Muons are identified either as a track reconstructed
in the muon chambers pointing back to the interaction vertex or by
their minimum-ionising-particle (MIP) signature in the central
tracking system and calorimeters.  Hadronic jets arising from $\tau$
decays are identified using a neural network based on topological
variables: the number of tracks and calorimetric clusters
associated to the jet, the half-opening angle of the jet, its
electro-magnetic energy and its mass.  The jet clustering algorithm
for a hadronically decaying $\tau$ is based on geometrical clustering
inside a cone of $15^\circ$ half-opening angle~\cite{GEOMJETS}. For
all other jets the Durham 
algorithm~\cite{DURHAM} is used.  The sum of the four-momenta of the
neutrinos in semileptonic, $\EEQQLNG$, and leptonic, $\EELNLNG$,
events is identified with the missing four-momentum of the event.

The following event generators are used to simulate the signal and
background processes: KORALW~\cite{KORALW}, HERWIG~\cite{HERWIG} and
ARIADNE~\cite{ARIADNE} ($\EEWWFFFFG$); EXCALIBUR~\cite{EXCALIBUR}
($\EEFFFFG$); PYTHIA~\cite{PYTHIA}  
($\EEQQG,\mathrm{ZZ}(\gamma),\mathrm{WW}(\gamma)$); 
KK2F~\cite{KK2F} ($\EEQQG$);
KORALZ~\cite{KORALZ}
($\EEMMG,~\TTG$); BHAGENE3~\cite{BHAGENE} and BHWIDE~\cite{BHWIDE}
($\EEEEG$); TEEGG~\cite{TEEGG} ($\EEEEGG$); DIAG36~\cite{DIAG36} (leptonic
two-photon collisions) and PHOJET~\cite{PHOJET}
(hadronic two-photon collisions).  The response of the L3 detector is
modelled with the GEANT~\cite{xsigel} detector simulation program
which includes the effects of energy loss, multiple scattering and
showering in the detector materials and in the beam pipe.  
Time dependent detector inefficiencies are taken into account in the simulation.
The selection efficiencies are
derived from Monte Carlo simulations fixing the
mass and the width of the W boson to $\MW=80.50\; \GeV$ and
$\GW=2.11\; \GeV$, respectively.

\subsection{\boldmath\protect\EELNLNG}
\label{sec:lnln}

The event selection for the process $\EELNLNG$ requires two high
energy acoplanar leptons with large missing energy due to the
neutrinos.  The selection depends on whether the event contains zero,
one or two identified electrons or muons, referred to as jet-jet,
lepton-jet and lepton-lepton classes, where jet denotes a hadronic
$\tau$-jet.  
The electron identification  is improved by
including also the SPACAL calorimeter that covers the polar
angular region between the barrel and the end-cap BGO calorimeters.  

\begin{figure}[p]
\begin{center}
{\epsfig{file=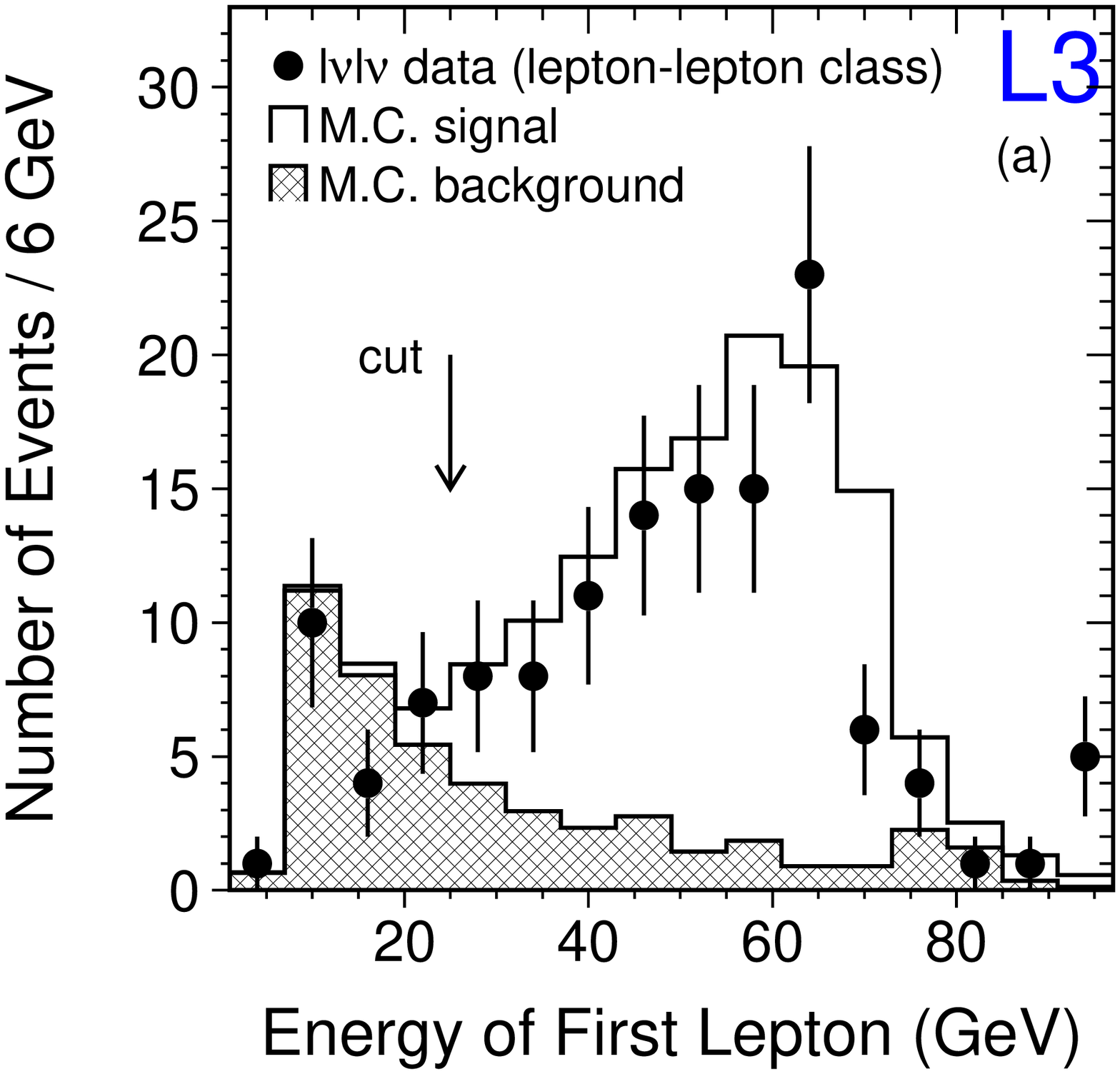,width=0.49\linewidth}}
{\epsfig{file=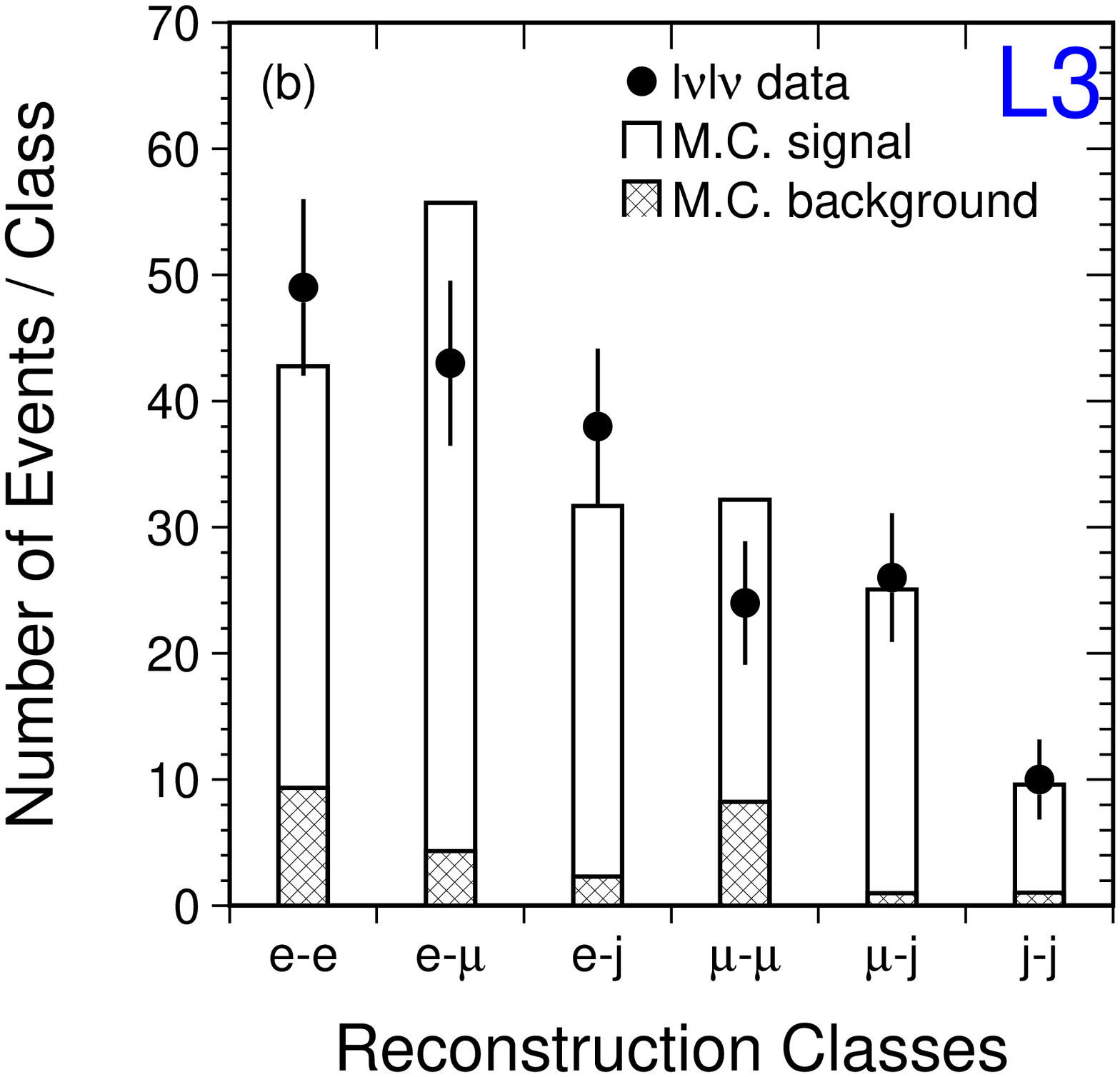,width=0.49\linewidth}}\\
{\epsfig{file=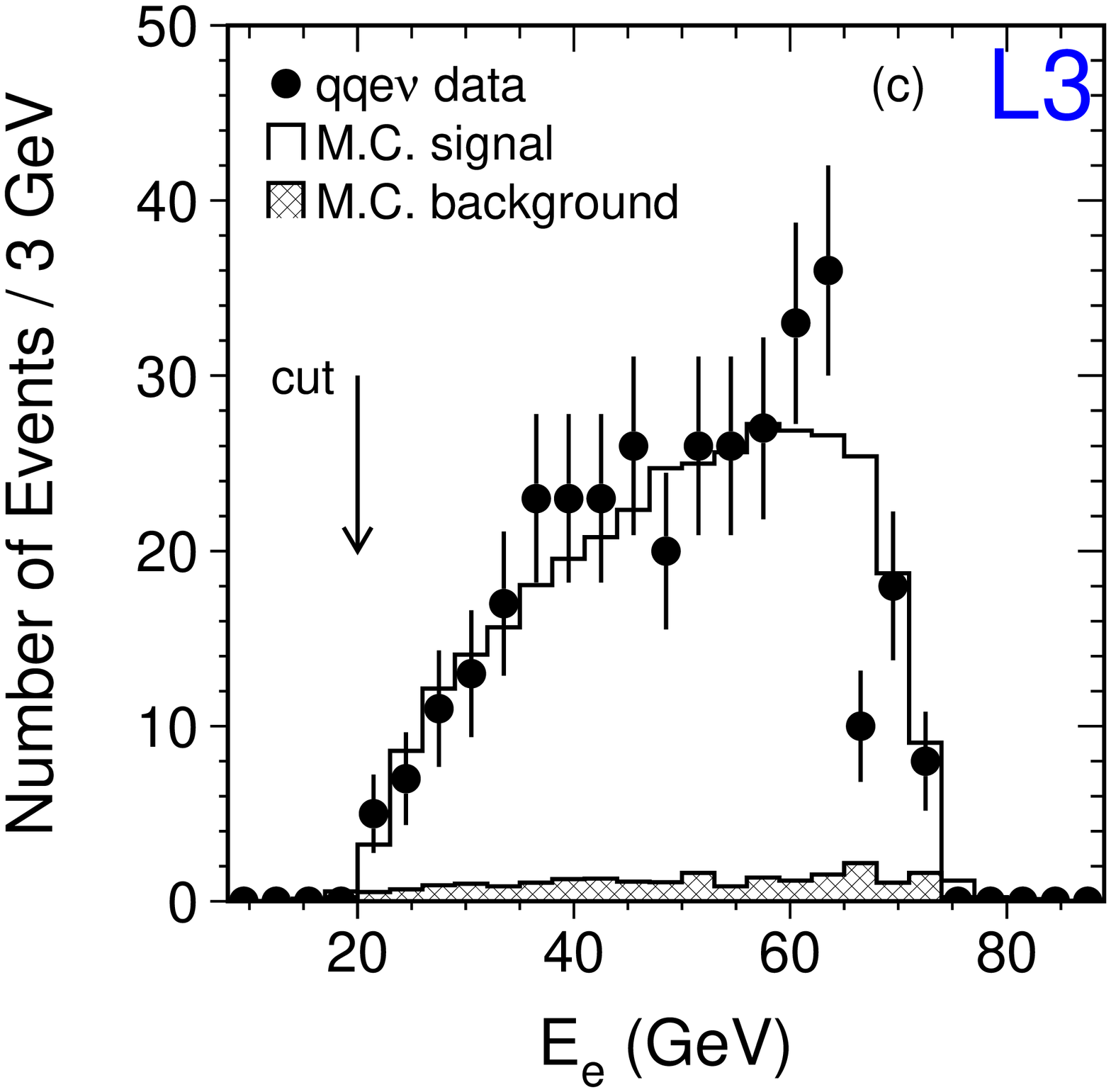,width=0.49\linewidth}}
{\epsfig{file=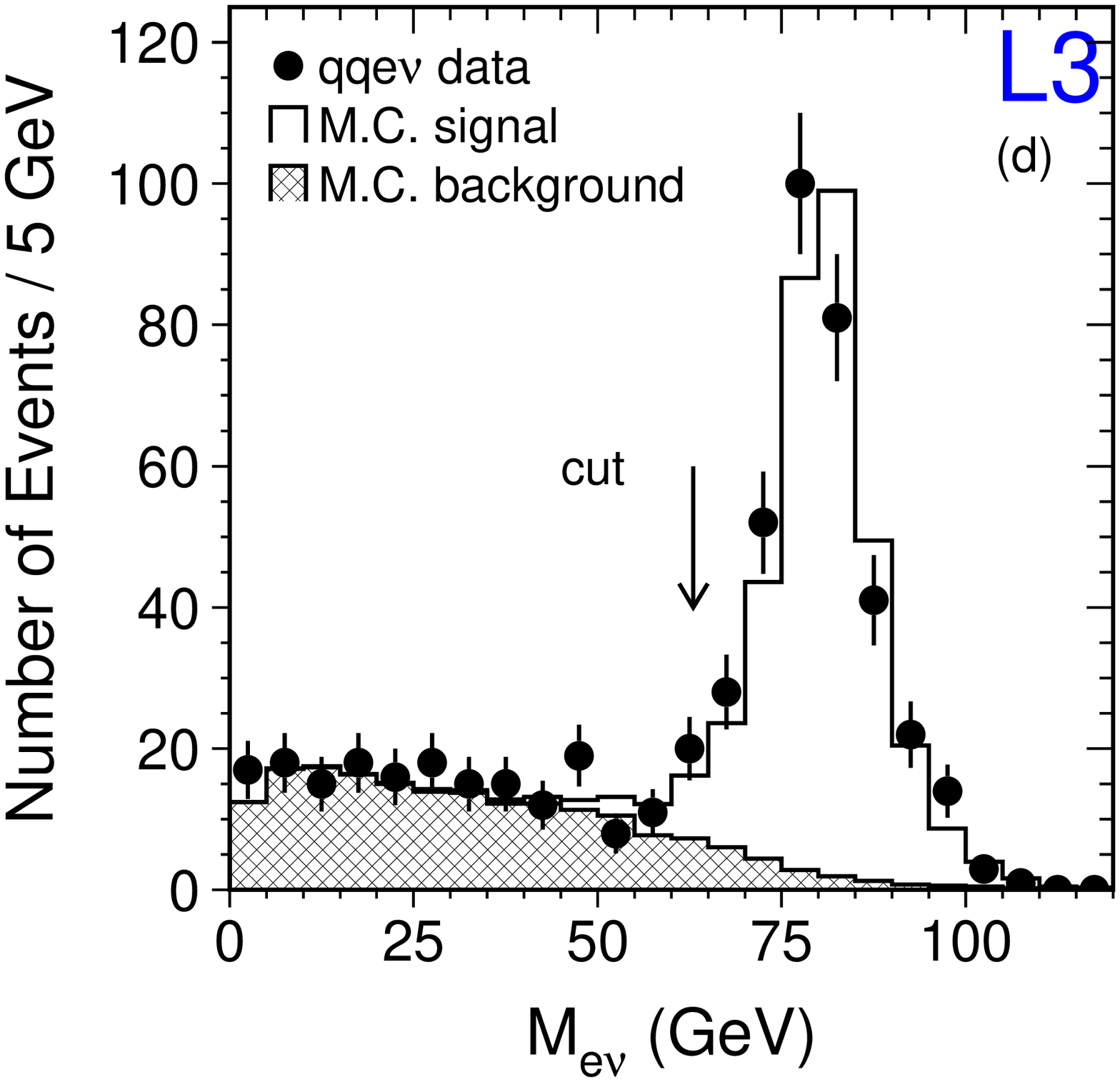,width=0.49\linewidth}}
\caption[]{Comparison of data and Monte Carlo distributions of
  variables used in the $\EELNLNG$ and $\EEQQENG$
  selections. The background consists of non-$\LNLN$/$\QQEN$ processes.
  (a) The energy of the most energetic electron or 
  muon in the lepton-lepton class. The
  vertical arrow indicates the cut position. The events in the plots
  have passed all other cuts. (b) The number of events in each final
  state topology (e=electron, $\mu$=muon, j= hadronic $\tau$-jet)
  after all cuts.
  (c) The energy, $\mathrm{E_e}$, of the electrons  
  identified in the BGO calorimeter.  (d) The
  invariant mass, $\mathrm{M_{e\nu}}$, of the electron-neutrino system.
}
\label{fig:lnlnqqen}
\end{center} 
\end{figure}

A total of 190 events is selected: 116 lepton-lepton
events, 64 lepton-jet events and 10 jet-jet events.  The distribution
of the energy of the highest-energy lepton in the lepton-lepton class
and the distribution of the selected events in the different reconstructed
final-state topologies are shown in Figure~\ref{fig:lnlnqqen}.

The signal efficiencies are determined with four-fermion (CC56+NC56)
Monte Carlo samples and are quoted for the following phase-space cuts:
$|\cos\theta|<0.96$ for both charged leptons, 
where $\theta$ is the polar angle with respect to
the beam axis, and energies greater
than $15~\GeV$ and $5~\GeV$ for the higher and the lower energy
lepton, respectively.  Table~\ref{tab:xmat-1} lists these efficiencies
in the form of a 6-by-6 matrix, relating $\LNLN$ events at generated
four-fermion level to $\LNLN$ events identified at reconstruction
level.  The overall selection efficiency in the full phase-space, 
based on a W-pair (CC03)
Monte Carlo and under the assumption of charged-current lepton
universality in W decays, is 52.3\%.
The background contributions, dominated by leptonic two-photon collisions
and $\EE \rightarrow \ell^+ \ell^- (\gamma)$ 
events, are listed in Table~\ref{tab:xmat-2}.

\subsection{\boldmath\protect\EEQQLNG}
\label{sec:qqln}

The selection of $\EEQQLNG$ events requires an
identified high energy lepton, two hadronic jets with high particle
multiplicity  and missing momentum due
to one or more neutrinos.  
The neutrino four-momentum vector is constructed by using the missing
three-momentum vector, taken to be massless. 
In order to separate $\QQENG$ and $\QQMNG$ events from $\QQTNG$
events for the case of a leptonic $\tau$ decay, the effective mass
of the lepton-neutrino system is used.
In the case of the $\WEN$ and $\WMN$ decay modes, the effective mass is
peaked around the W mass, whereas for $\WTN$ decays its value is
typically much lower.
The cut position is chosen such that the correlation
among the cross sections of the three final states is minimised.  
The
selection efficiencies and the background
contaminations, dominated by $\QQG$ events, 
are presented in 
Tables~\ref{tab:xmat-1} and~\ref{tab:xmat-2}.

The $\QQENG$ selection accepts 363 events.
The distributions of the energy of the electron and of the
electron-neutrino invariant mass are shown in Figure~\ref{fig:lnlnqqen}.
The signal efficiency for $\QQENG$ events is determined from a
four-fermion (CC20) Monte Carlo sample within the following
phase-space cuts: $E_{\e},E_\nu>15~\GeV$, where $E_{\e}$ and $E_\nu$
are the electron and neutrino energies; 
$|\cos\theta_{\e}|,|\cos\theta_\nu|<0.98$, where $\theta_{\e}$ and
$\theta_{\nu}$ are the electron and neutrino polar angles; 
$M_{\e\nu},M_\mathrm{qq}>45~\GeV$, where $M_{\e\nu}$ and $M_\mathrm{qq}$ are
the electron-neutrino and quark-quark invariant masses.
These values are changed with respect to our previous analysis due to
the increased centre-of-mass energy.

\begin{figure}[p]
\begin{center}
{\epsfig{file=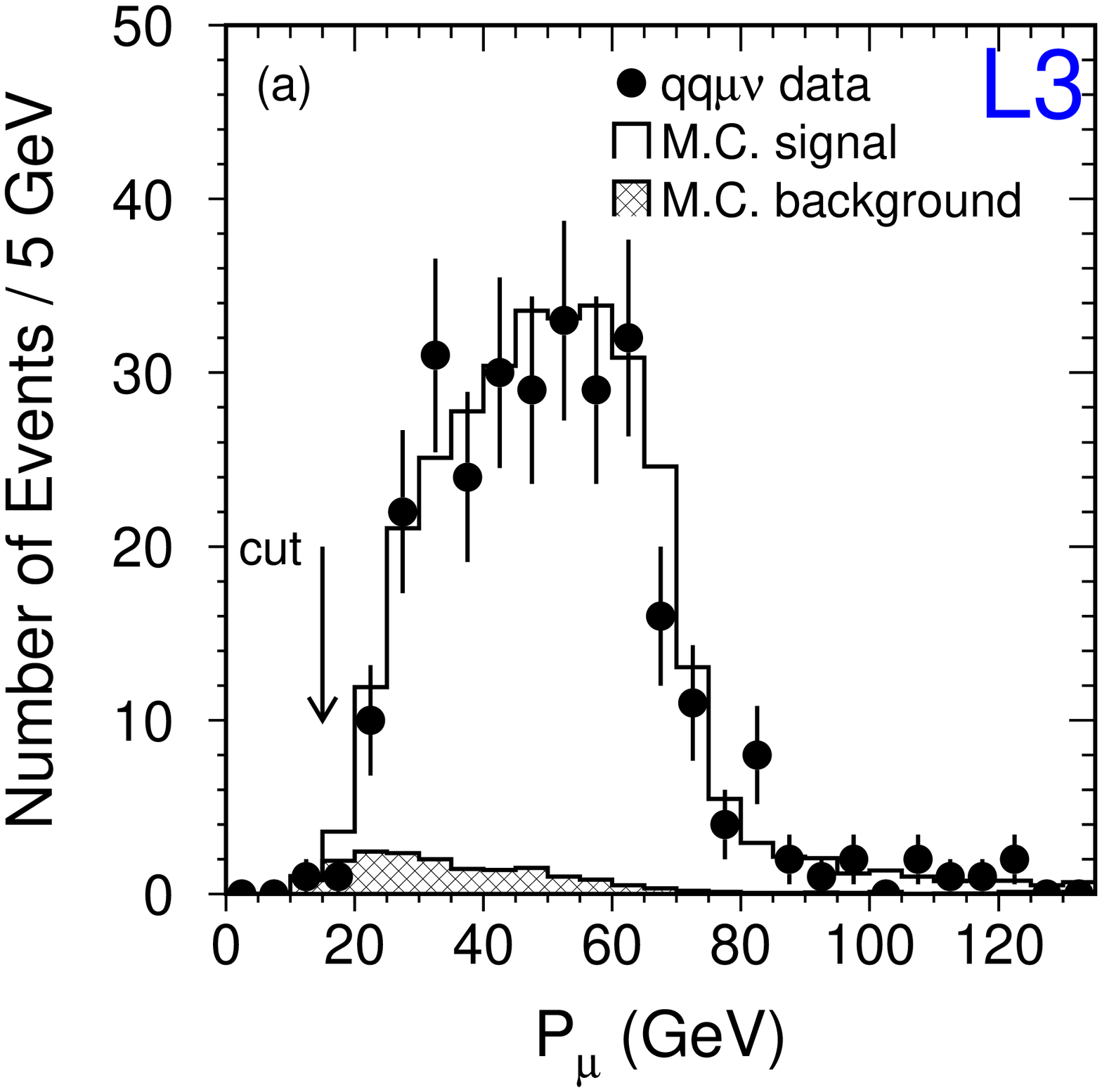,width=0.49\linewidth}}
{\epsfig{file=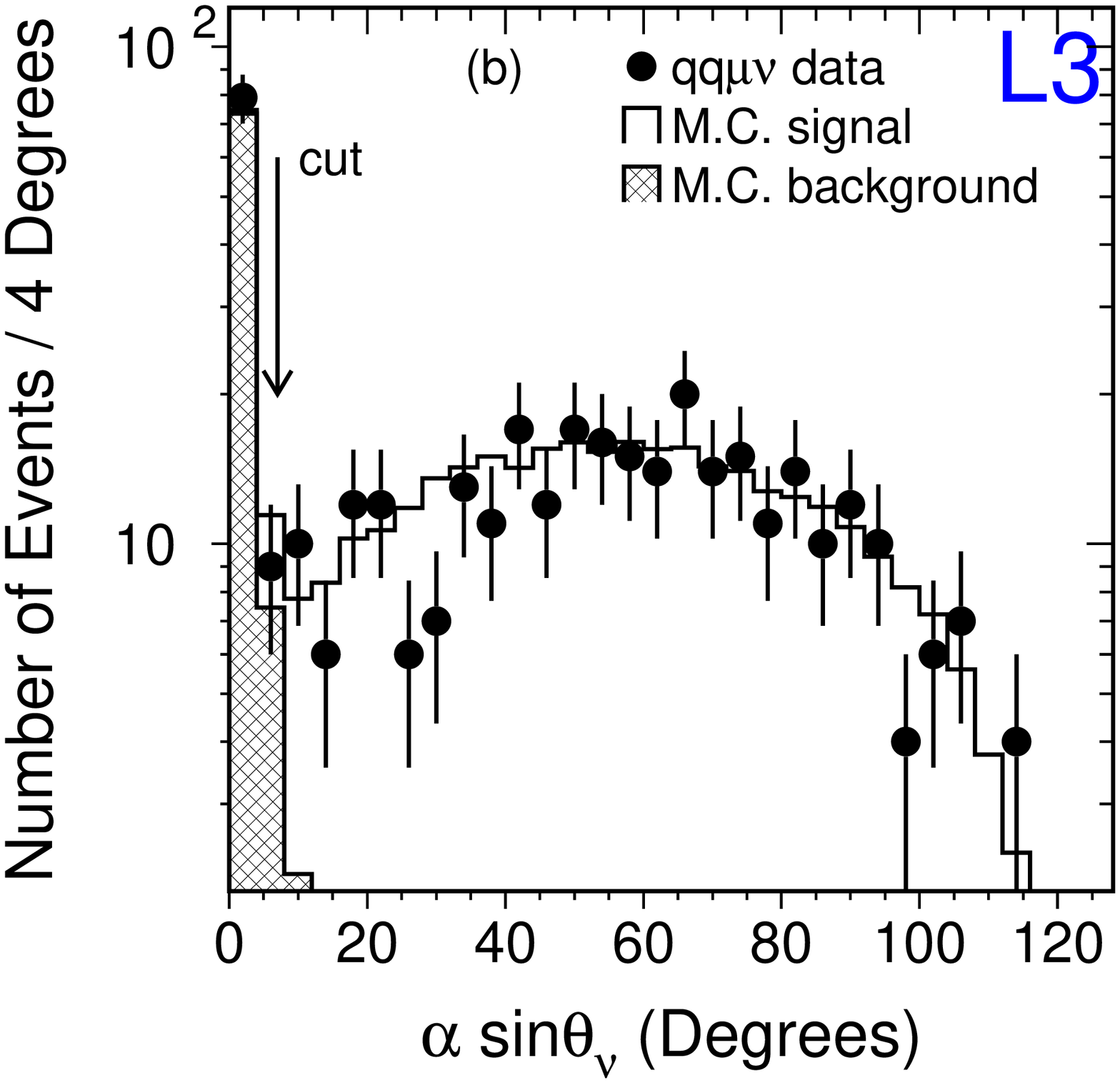,width=0.49\linewidth}}\\
{\epsfig{file=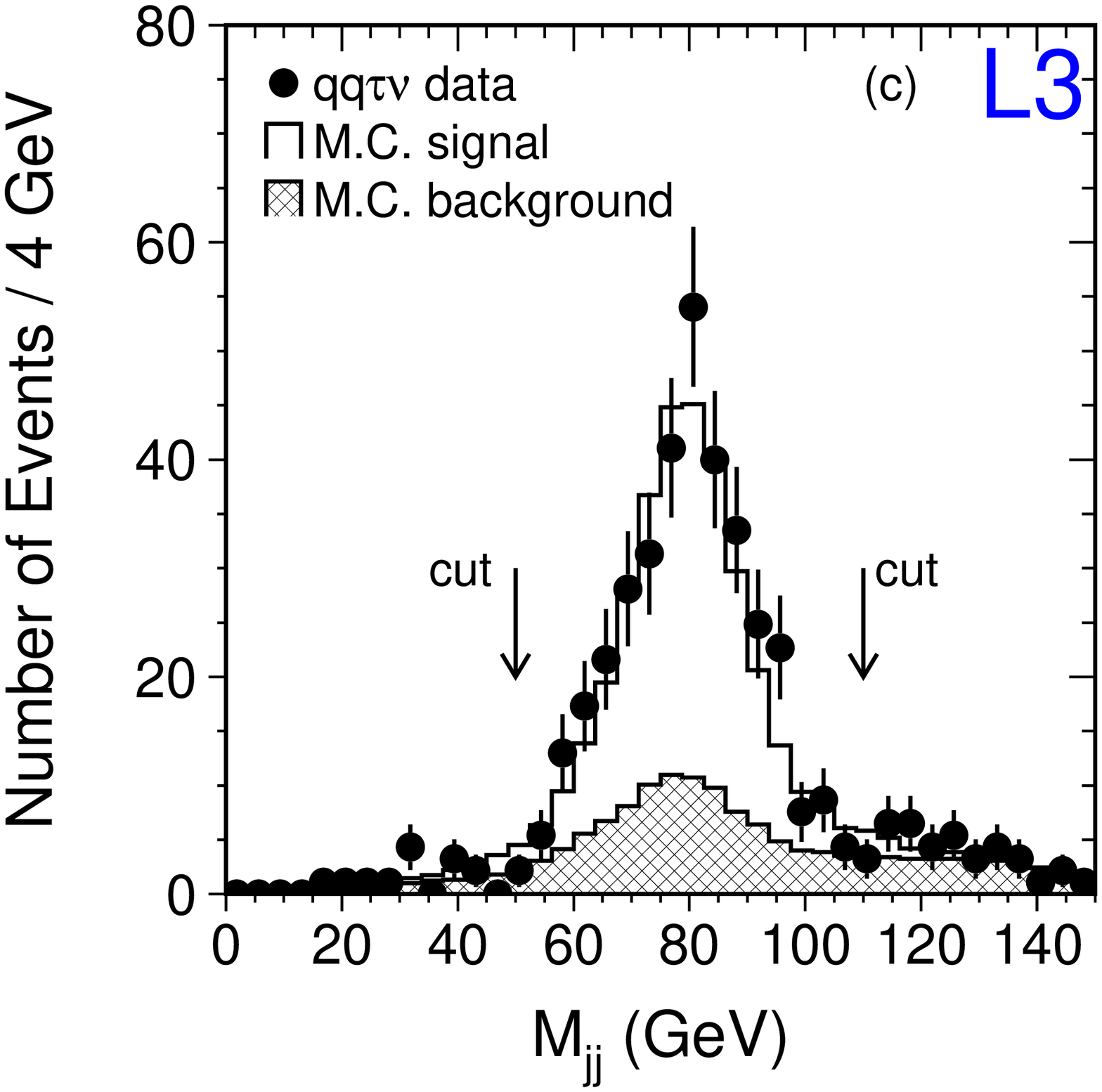,width=0.49\linewidth}}
{\epsfig{file=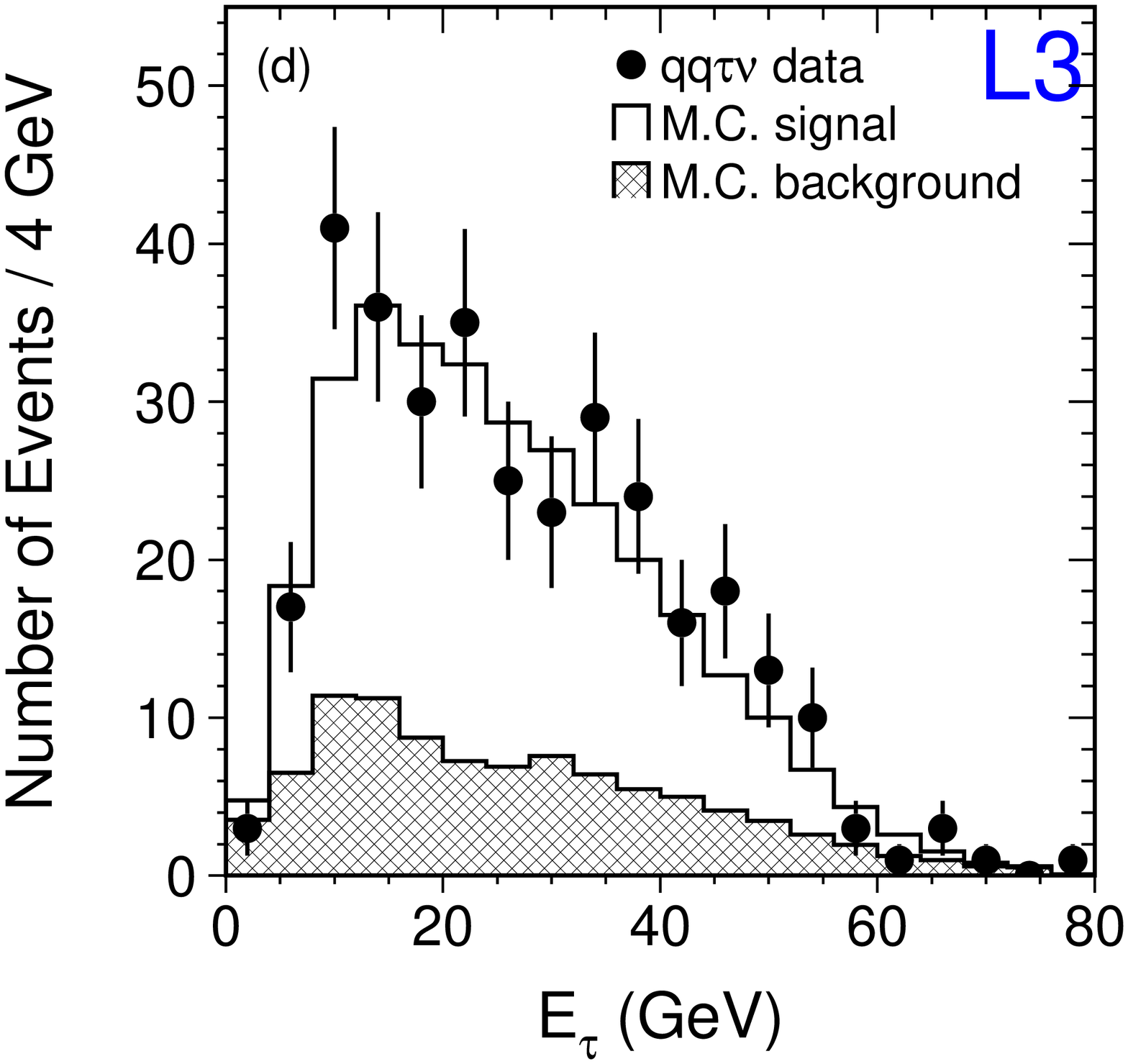,width=0.49\linewidth}}
\caption[]{Comparison of data and Monte Carlo distributions of
  variables used in the $\EEQQMNG$ and $\EEQQTNG$
  selections. 
  (a) The momentum, $\mathrm{P_\mu}$, of the muons reconstructed 
      in the muon chambers.
  (b) The variable $\alpha \sin \theta_\nu$ described in the text.
  (c) The invariant mass, $\mathrm{M_{jj}}$, of the jet-jet system.
  (d) The energy, $\mathrm{E_\tau}$, of the visible decay products of 
      the $\tau$.}
\label{fig:qqmnqqtn}
\end{center} 
\end{figure}

A total of 340 events are accepted by the $\QQMNG$ selection:
299 candidates with a muon reconstructed in the muon spectrometer and 41
with a muon identified by its MIP signature.  The distributions
of the momentum of the muon and of the variable $\alpha
\sin \theta_\nu$, which discriminates against the $\QQG$
background events, are presented in Figure~\ref{fig:qqmnqqtn}.
The angle $\alpha$ is that between the direction of the muon and the
nearest hadronic jet and $\theta_\nu$ is the polar angle of the
neutrino.

A total of 329 $\QQTNG$ events are selected: 53
$\tau\rightarrow\nu\e\bar\nu$, 50
$\tau\rightarrow\nu\mu\bar\nu$ and 226
$\tau\rightarrow\nu+$hadrons candidates.  The distributions  of the jet-jet 
invariant mass and of the visible $\tau$ energy are shown in
Figure~\ref{fig:qqmnqqtn}.

\subsection{\boldmath\protect\EEQQQQG}
\label{sec:qqqq}
%
%
The $\EEQQQQG$ selection requires high multiplicity four-jet events
with low missing energy.
It accepts 94.8\% of the $\WW\rightarrow\QQQQG$ signal 
with a purity of 48.4\%, corresponding to 
2674 selected events.  The measurement of jet
energies and angles is improved by a kinematic fit that imposes 
four-momentum conservation.

\begin{figure}[p]
\begin{center}
{\epsfig{file=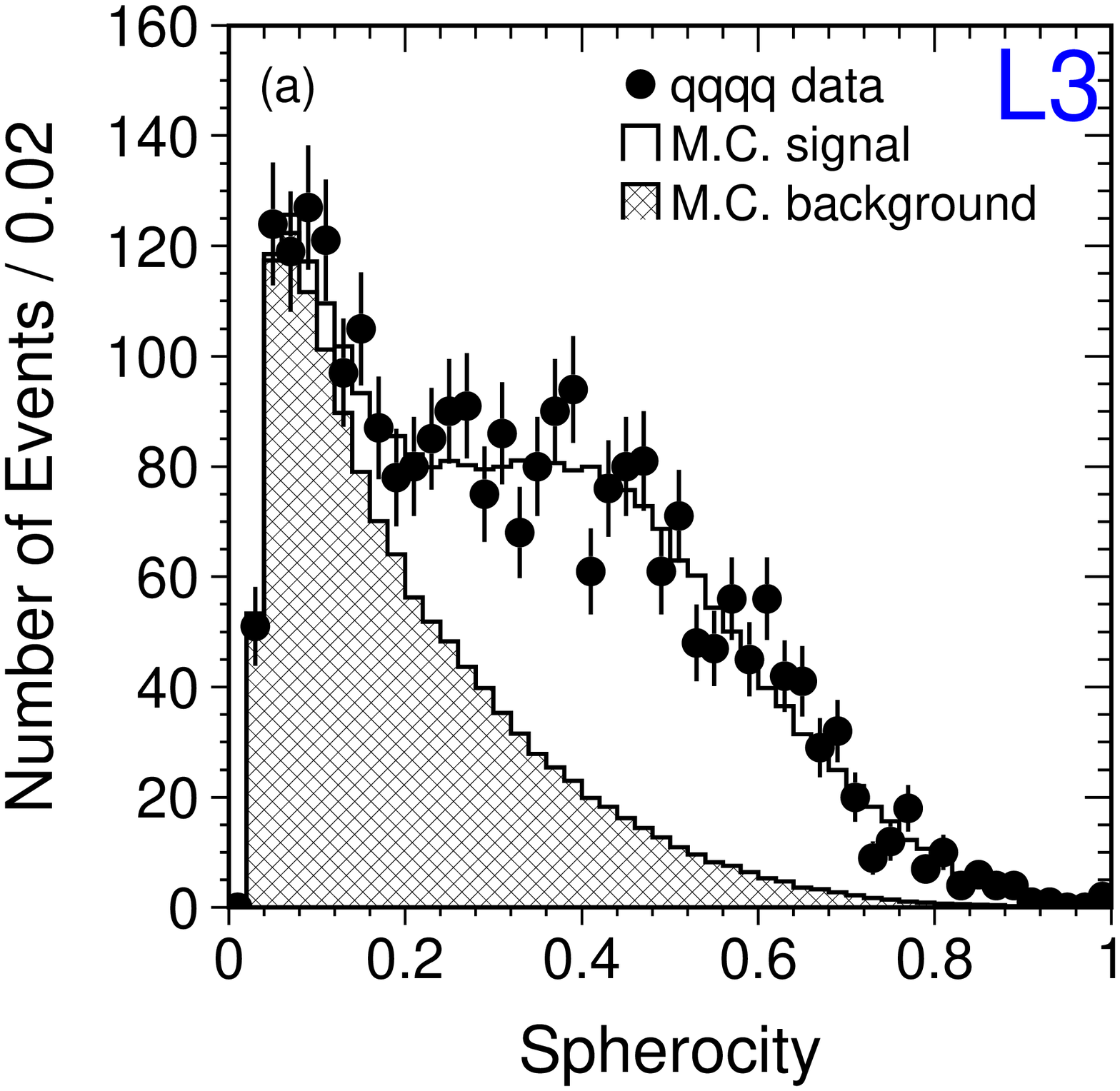,width=0.49\linewidth}}
{\epsfig{file=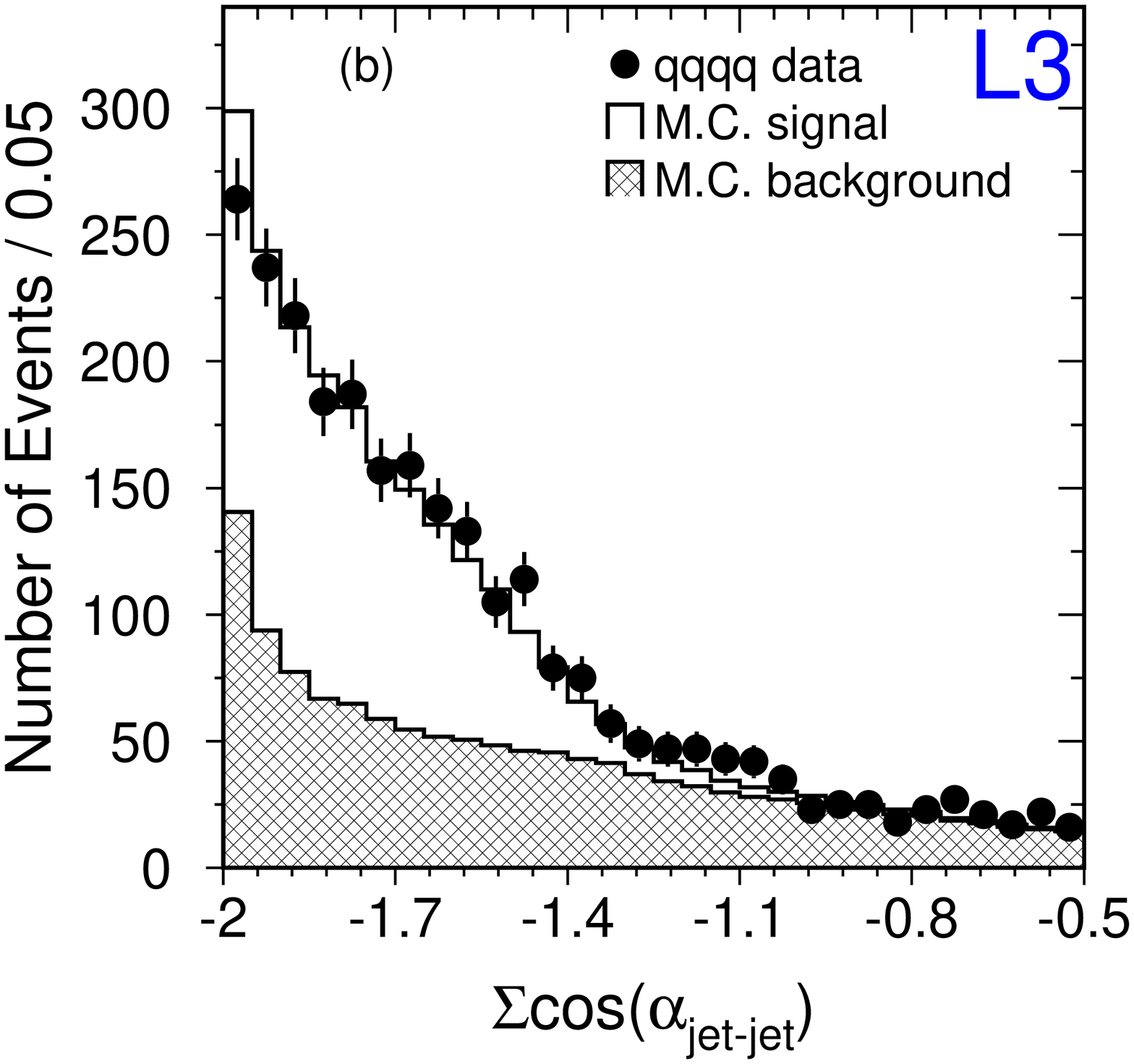,width=0.49\linewidth}}\\
{\epsfig{file=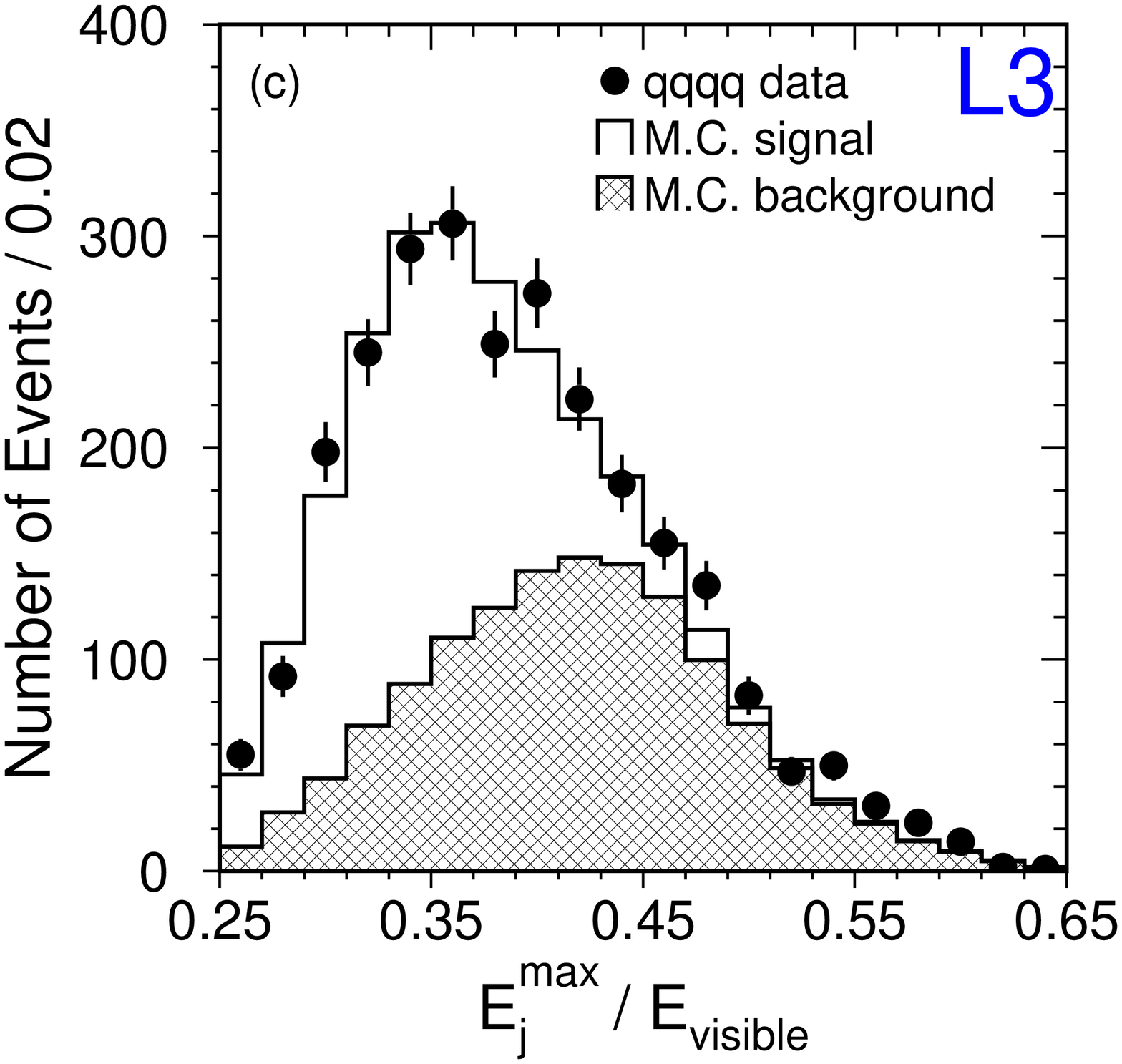,width=0.49\linewidth}}
{\epsfig{file=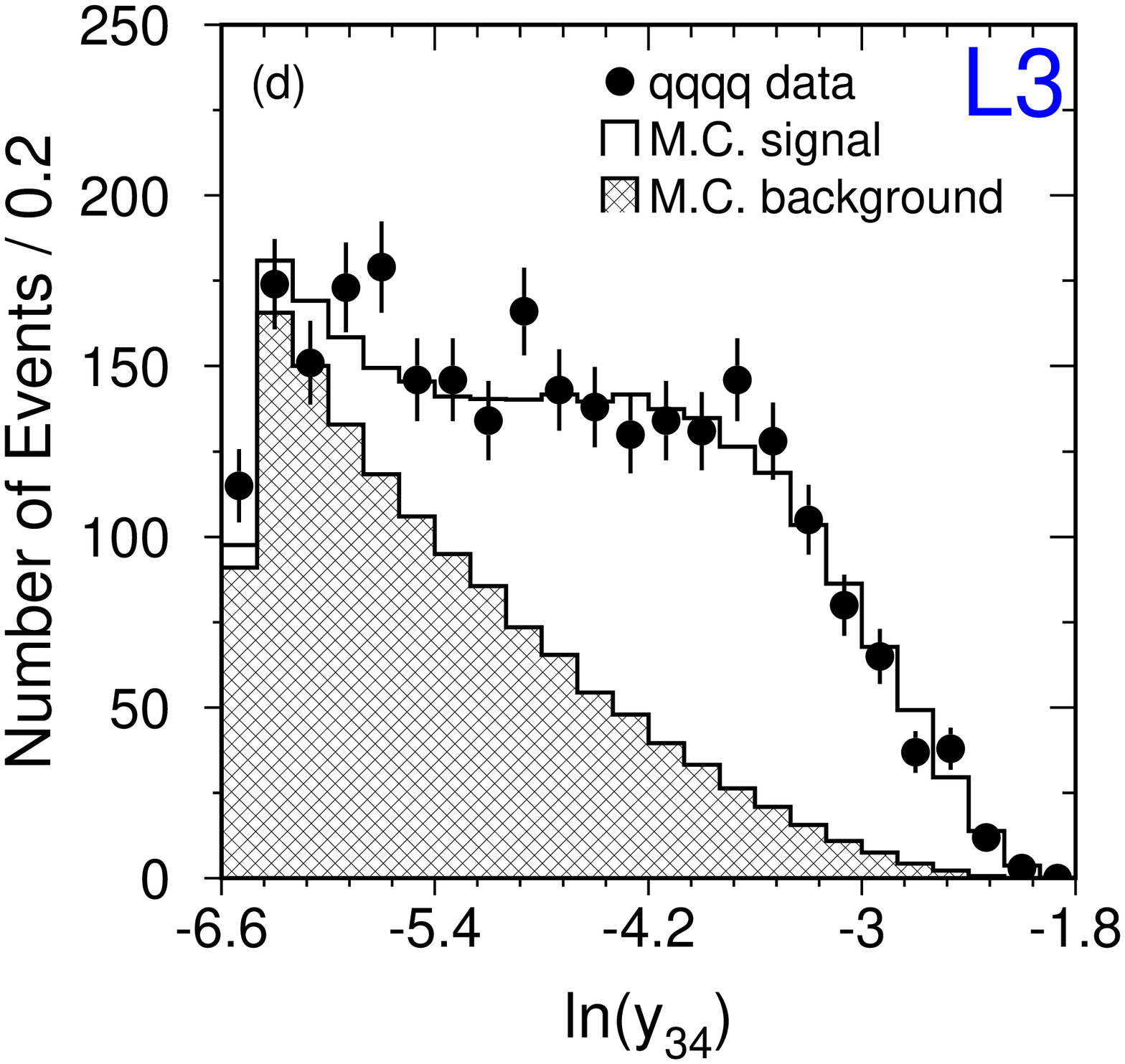,width=0.49\linewidth}}
\caption[]{
  Distributions of input variables for the neural network used in the
  analysis of the $\EEQQQQG$ process.
  All selection cuts are applied.
  (a) The spherocity.
  (b) The sum of cosines of the jet-jet angles.
  (c) The maximal jet energy, $\mathrm{E_j^{max}}$, 
      normalised to the total visible energy.
  (d) The quantity $\ln(y_{34})$.
}
\label{fig:qqqq-nnin}
\end{center}
\end{figure}

\begin{figure}[p]
\begin{center}
{\epsfig{file={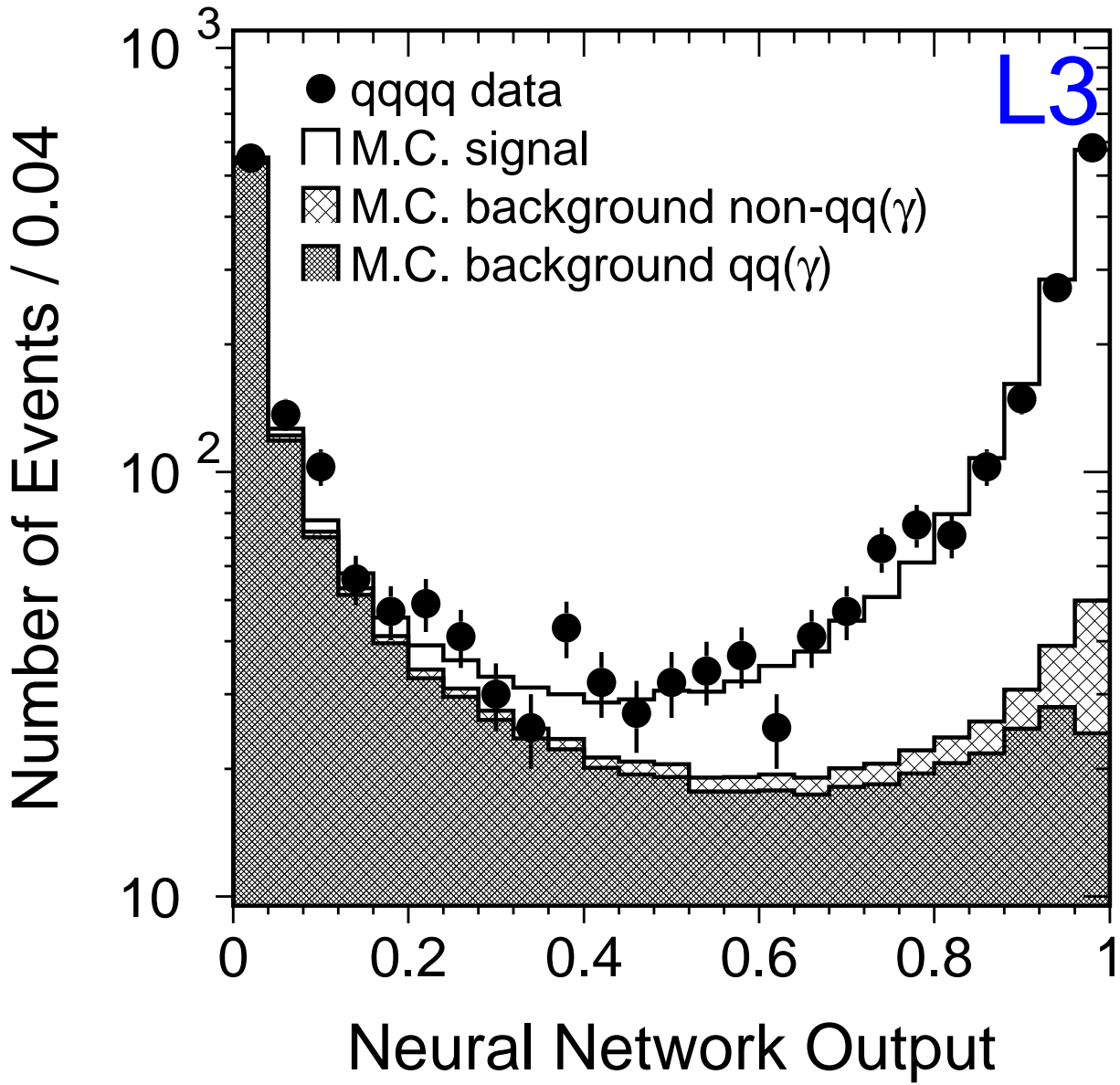},width=\linewidth}}
\caption[]{Comparison of data and Monte Carlo distributions of
  the neural network output for selected $\EEQQQQG$ events.  
}
\label{fig:qqqq-nnout}
\end{center} 
\end{figure}

A neural network is trained to further separate the signal from the main
$\EEQQG$ background, which is actually dominated by 
events with multi-gluon radiation.  
The input to the network consists of ten event
variables: minimal and maximal jet energy, the energy difference
between the two remaining jets, the minimal jet cluster multiplicity,
the logarithm of the Durham jet-resolution parameter $y_{34}$  
at which the event
changes from a four-jet to a three-jet topology, 
the spherocity~\cite{LEP1YRSP}, the sum of the cosines of
the jet-jet angles, the probability of the kinematic fit and the jet
broadening of the most and least energetic jets. The jet broadening is
defined as $\sum \sqrt{p_t}/\sum \sqrt{p}$ where the sum
is over the particles belonging to the jet, $p_t$ is the transverse
momentum relative to the reconstructed jet axis and $p$ is the
momentum of the particle.  
The distributions of
four of these variables are shown in Figure~\ref{fig:qqqq-nnin}.  The
network is trained such that its output peaks at one for the signal
and at zero for the background, as shown in
Figure~\ref{fig:qqqq-nnout}.

The neural network output distribution for data events is fitted by a
linear combination of outputs derived from
Monte Carlo simulations for signal and background. A binned maximum
likelihood fit is performed leaving both the signal  and
the most significant background ($\EEQQG$) cross sections as free
parameters.
The latter is measured as $104.9\pm3.4$~pb, 
in agreement with both our measurement~\cite{l3-196} 
and the SM value of $96.9$~pb.

\section{Systematic Uncertainties}
\label{sec:syst}

The systematic uncertainties taken into account in the determination of
the results are summarised in Table~\ref{tab:syst}.  Common to all
final states is the uncertainty on the integrated luminosity 
of 0.2\%~\cite{l3-196}.  
The Monte Carlo statistics for both signal and background processes
affects mainly the $\LNLNG$ final states.  

The uncertainty due to the selection procedure 
is evaluated by varying
the positions of the selection cuts and interpreting  the corresponding
changes in the measured cross sections as systematic uncertainties.  
The systematic uncertainty on the neural network output
in the $\QQQQG$ selection is estimated by recalculating the input
variables of the neural network after smearing and scaling the
measurements of energy depositions and tracks in the simulation 
according to the uncertainty on their resolutions.
In all selections, 
consistent results are found by studying the change in efficiency due
to variations of 
the detector calibration within its uncertainty.  The calibration is studied
using samples of di-lepton and di-jet events,
collected during the calibration runs at $\sqrt{s}=91~\GeV$ and at higher
energies.
The trigger inefficiency as well as its uncertainty is found to be
negligible.  
The systematic uncertainty assigned to the selection procedure is
in the range 1.1\% to 2.5\% depending on the final state.

The theoretical uncertainties on the background cross sections, mainly on
hadronic two-photon collisions (50\%) and neutral-current four-fermion
processes (5\%),
are propagated to the W-pair cross sections.
The observed change of 0.1\% to 0.3\% is assigned as 
a systematic uncertainty.  The KK2F Monte
Carlo generator, instead of PYTHIA, is also used to simulate the
$\QQG$ background revealing no significant deviation on the measured
cross section.

The dependence of the selection efficiencies on the mass and width of
the W boson is studied using Monte Carlo samples simulated with
different $\MW$ and $\GW$ values. The propagation of the world average
uncertainties on these two parameters~\cite{PDG98}, 100~$\MeV$ on $\MW$ and
60~$\MeV$ on $\GW$, is taken as a systematic uncertainty and  ranges from
0.1\% to 0.3\%.

The systematic uncertainty on initial-state radiation (ISR) due to its
approximate (leading-log) treatment in KORALW  is
investigated by reweighting the energy and transverse momentum spectra
of the ISR photons. This is done according to an exact ${\cal O}(\alpha)$
matrix element calculation~\cite{ISR}
that also includes leading ${\cal O}(\alpha^2)$
contributions at small angles. This study shows that the
associated systematic uncertainty on the cross section determination
is negligible. 
For the $\LNLNG$ and $\QQENG$ selections, the efficiencies
are determined by using EXCALIBUR Monte Carlo events. 
To estimate the systematic uncertainty due to the
restriction in the EXCALIBUR program to strictly collinear ISR
photons, the KORALW generator, which includes both collinear and
non-collinear ISR photons, is used. The change in the selection
efficiency of KORALW events given by including or removing the
non-collinear contribution is used as an estimate of the similar
effect of neglecting such photons in EXCALIBUR.  
The $\LNLNG$ selection, requiring low activity in the regions of
the detector away from the leptons, is  
sensitive to the transverse momentum of radiated photons.
The resulting reduction of 0.5\% absolute is applied as a correction to
the EXCALIBUR efficiency and it is
included in the results reported in Table~\ref{tab:xmat-1}.
For the $\QQENG$ final state, the correction turns out to be negligible.
The total uncertainty on the
correction  in the $\LNLNG$ case, propagated to the 
cross section, amounts to 0.2\% and is assigned 
as the systematic uncertainty due to ISR.

Final-state radiation (FSR) is absent in EXCALIBUR while it is
implemented in  KORALW by using the PHOTOS package~\cite{PHOTOS} based
on the leading-log approximation. 
A similar procedure as for ISR is then applied using the KORALW
program to correct the efficiencies for the complete absence of 
FSR in EXCALIBUR in the case of  the $\LNLNG$ and $\QQENG$ selections.
The correction amounts to a reduction in efficiency of
0.4\% absolute for the $\LNLNG$ and 1.3\% absolute for the 
$\QQENG$ final states
and is included in the results listed in Table~\ref{tab:xmat-1}. 
The effect on the $\QQENG$ selection comes from its sensitivity to the
electron isolation requirements affected by FSR photons. The
PHOTOS package is inaccurate in the hard non-collinear region~\cite{PHOTOS}.
The related systematic
uncertainty is estimated by determining  the selection efficiencies 
using Monte Carlo signal events containing soft radiative photons
only. 
Half of the difference with respect to the full radiation simulation, 
ranging from 0.2\% to 0.9\%, 
is taken as a systematic uncertainty.

The determination of the efficiency for the $\QQMNG$ and $\QQTNG$
selections, as well as the neural network output shape for the $\QQQQG$
channel, is based on a W-pair (CC03) Monte Carlo sample.  This Monte
Carlo neglects non-WW contributions interfering with the CC03
diagrams.  In order to study the effect on the cross section,
the CC03 Monte Carlo events are reweighted 
by the squared matrix element ratio of the
full four-fermion calculation to the CC03 one.
In the case of $\QQQQG$ selection, this reweighting procedure takes
into account that the $\mathrm{ZZ}(\gamma)$ final states are treated
in the background subtraction.
The difference in accepted cross section, in the range 0.1\% to 0.4\%,
is taken as a systematic uncertainty.

The effect of modelling of the signal hadronisation is studied comparing the
selection efficiencies based on simulated samples with different
hadronisation schemes, JETSET~\cite{PYTHIA}, HERWIG~\cite{HERWIG} and
ARIADNE~\cite{ARIADNE}.  The hadronisation parameters in these models
were tuned to describe inclusive hadronic Z decays at
$\sqrt{s}=91~\GeV$.  No difference is observed for the $\QQENG$ and
$\QQMNG$ cross sections and the statistical accuracy of the test,
0.3\%, is taken as a systematic uncertainty.  In the case of the
$\QQTNG$ selection, the treatment of hadronic jets and the tight
requirements needed in order to identify the $\tau$ jet,
lead to an uncertainty of 1.1\%.  In the $\QQQQG$ case a larger
difference of 1.8\%, dominated by the HERWIG comparison, 
is observed and assigned as the signal hadronisation
uncertainty.

The hadronisation systematics due to the $\EEQQG$ background is
negligible for $\QQLN$ final states. For the $\QQQQG$ selection it is
estimated by two methods. In the first, a sample of HERWIG $\QQG$
events is used instead of JETSET events.  In the second, data and
Monte Carlo distributions of hadronic Z decays collected at
$\sqrt{s}=91~\GeV$ are compared for the $y_{34}$ variable.  The
discrepancy is treated by reweighting Monte Carlo events to
match the data as a function of $y_{34}$.  The resulting accepted
$\QQG$ cross section for a neural network output greater than 0.6 is
increased by 4.8\% relative to the JETSET prediction, while the signal
cross section is decreased by 0.6\%.  These corrections are
applied to the quoted results.  From the uncertainty on the reweighting
procedure and from the HERWIG$-$JETSET comparison, a 0.6\% systematic
uncertainty is assigned.

The modelling of Bose-Einstein correlations between hadrons from W
decays may affect the selection efficiencies.  
In our recent study~\cite{l3-213} we have measured the
strength of intra-W Bose-Einstein correlations in semileptonic W decays. Its
value is significantly different from zero and in good agreement with
that for light-quark Z decays and with 
the LUBOEI~\cite{LUBOEI} BE$_{32}$ and BE$_{0}$ predictions.
The resulting systematic uncertainty is found to be negligible.
  
The same study shows that Bose-Einstein
correlations between particles originating from different W bosons are
strongly disfavoured in $\QQQQG$ events. Their strength is restricted to
at most 1/4  of the strength as simulated in the BE$_0$/BE$_{32}$
models with full correlations.  The difference in the measured
$\QQQQG$ cross section obtained by using Monte Carlo samples with
intra-W only and with full inter-W correlations is rescaled by 1/4 according
to the allowed strength of inter-W correlations. 
The resulting difference of 0.3\%
is assigned as a systematic uncertainty. The uncertainty coming from the
modelling of intra-W correlations is negligible as for the $\QQLNG$ case.

The influence of colour reconnection effects between the two hadronic
systems arising in $\WW\rightarrow\QQQQG$ decays  are estimated using
models implemented in ARIADNE (model 2)
and PYTHIA 6.1 (models~\cite{SKMODELS} SK I with reconnection
parameter $k=0.6$, SK II and SK II$^\prime$).  In all cases, the
effect on the cross section is at the level of the statistical accuracy
of 0.3\%, which is then taken as the associated systematic uncertainty.

The total systematic uncertainty on the cross sections is 
estimated to be in the range 1.5\%$-$2.9\%, depending on the final state, 
from the combination of the above considered sources.  

\section{Results}
\label{sec:results}

The CC03 cross sections $\sigma_i$ of the ten signal processes $i$, or
equivalently the W-decay branching fractions together with the total
W-pair cross section, are determined simultaneously in
maximum-likelihood fits~\cite{L-XSEC-161-183}. 
For the $\QQENG$ and the
six $\LNLNG$ final states, where the non-W-pair processes have a
significant contribution, the efficiencies are determined within the
phase-space cuts described in Section~\ref{sec:lnln}
and~\ref{sec:qqln}.  The measured
four-fermion cross
sections are scaled by the conversion factors $f_i$, listed in
Table~\ref{tab:xsec}, to obtain the CC03 cross sections.
These factors are given by the ratio of the total CC03 cross
section and the four-fermion cross section within phase-space cuts and
are calculated with the EXCALIBUR
Monte Carlo program.

The SM CC03 predictions are obtained using the GENTLE~\cite{GENTLE} 
program and also the new calculations implemented in the RacoonWW~\cite{RACONWW}
and YFSWW3~\cite{YFSWW3} programs that include full ${\cal
  O}(\alpha)$ electroweak corrections, calculated in the double
pole approximation~\cite{DPA}. The new predictions are about
2.5\% lower than the GENTLE calculations, and have a reduced theoretical
uncertainty of about 0.5\% instead of 2\%~\cite{LEP2MC4F}. For numerical comparisons,
RacoonWW predictions are used.

\subsection{Signal Cross Sections}

The ten measured 
CC03 signal cross sections $\sigma_i$ are listed in Table~\ref{tab:xsec}.
They agree well with the SM expectations.  
The correlations between the measured cross sections, reported in
Table~\ref{tab:corrmatrix}, arise from non-vanishing off-diagonal
terms in the efficiency matrix of Table~\ref{tab:xmat-1}.  The
systematic uncertainties shown in Table~\ref{tab:syst} and their
correlations are taken into account in the determination of the
results.

With the assumption of charged current lepton universality in W
decays, the number of free parameters in the fit is reduced to three CC03
cross sections: $\sigma_{\LNLN}$, $\sigma_{\QQLN}$, 
summed over all lepton species, and
$\sigma_{\QQQQ}$. Their values at
$\sqrt{s}=188.6~\GeV$ are determined to be:
\begin{eqnarray}
\sigma_{\LNLN} & = & 1.67\pm0.14\pm0.04~\mathrm{pb}\nonumber\\
\sigma_{\QQLN} & = & 7.20\pm0.24\pm0.08~\mathrm{pb}\nonumber\\
\sigma_{\QQQQ} & = & 7.36\pm0.24\pm0.18~\mathrm{pb}\,,
\end{eqnarray}
where the first uncertainty is statistical and the second systematic.  
The correlations between these three results are negligible.

\subsection{W-Decay Branching Fractions and W-Pair Cross Section}

The W decay branching fractions $B(\WFF)$ are derived in a combined
fit with the total CC03 W-pair cross section. 
The sum of the branching fractions is fixed to be unity and their values 
are combined with our measurements at lower centre-of-mass
energies~\cite{L-XSEC-161-183} including correlations due to
systematic uncertainties. The results
are presented in Table~\ref{tab:brafra}, both without and with the
assumption of charged current lepton universality in W decays. 
This hypothesis is supported by our results.
The values for the assumption of lepton universality are:
\begin{equation}
B(\WQQ) =  (68.20\pm0.68\pm0.33)~\% \qquad
B(\WLN) = (10.60\pm0.23\pm0.11)~\% \,.
\end{equation}
The relation between the W-decay branching fractions and the six
elements $V_{ij}$ of the Cabibbo-Kobayashi-Maskawa quark mixing matrix
$V_{\mathrm{CKM}}$~\cite{VCKM} not involving the top
quark is~\cite{LEP2YRWW}:
$1/B(\WLN)=3+3\left(1+\alpha_{\mathrm{s}}(\MW)/\pi\right)V^2$.
Using $\alpha_{\mathrm{s}}(\MW)=0.121 \pm0.002$ as the strong coupling
constant, our measurement yields:
\begin{eqnarray}
V^2 & = & \sum_{i=\mathrm{u,c};\,j=\mathrm{d,s,b}}
  \left|V_{ij}\right|^2 
    ~ = ~ 2.065 \pm 0.064 \pm 0.032 \,.
\end{eqnarray}
Using the current world-average values and uncertainties of the other
matrix elements~\cite{PDG98},
the value of  $V_{\mathrm{cs}}$ is derived as:   
\begin{eqnarray}
  |V_{\mathrm{cs}}| & = & 1.008 \pm 0.032 \pm 0.016 \,.
\end{eqnarray}
This element is the least known of the
two dominant diagonal elements appearing in $V^2$.
The statistical uncertainty includes the uncertainties on
$\alpha_{\mathrm{s}}$ and the other $V_{ij}$ elements but is dominated
by the statistical uncertainty on our measured W-decay branching
fractions.
 
\begin{figure}[p]
\begin{center}
{\epsfig{file=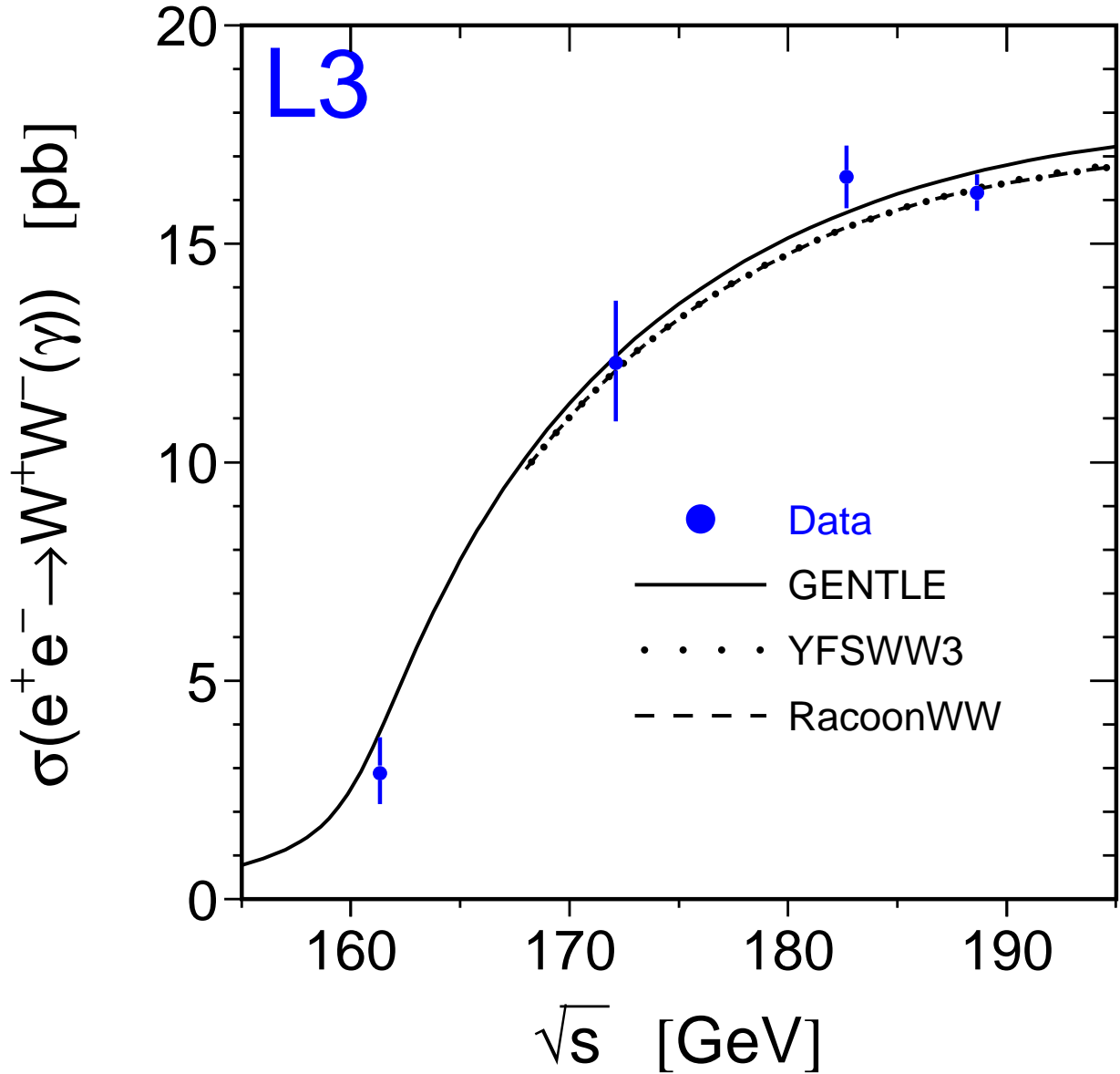,width=\linewidth}}
\caption[]{
  The CC03 cross section of the process $\EEWWG$ as a function
  of the centre-of-mass energy.  The published
  measurements at $\sqrt{s}=161$, $172$  
  and $183~\GeV$, and the new measurement at
  $\sqrt{s}=189~\GeV$ are shown combining
  statistical and systematic uncertainties in quadrature.  The 
  curves show the SM expectation according to different calculations.
}
\label{fig:xsec}
\end{center}
\end{figure}

Assuming SM W-decay branching fractions~\cite{LEP2YRWW}, the total
W-pair cross section at $\sqrt{s}=188.6~\GeV$ is measured to be:
\begin{eqnarray}
  \SWW & = &  16.24 \pm 0.37 \pm 0.22~\mathrm{pb}\,.
\label{eq:xsec-ww}
\end{eqnarray}
The measurements reported here are in agreement with recent measurements by
other LEP experiments at the same centre-of-mass
energy~\cite{ADXSEC-189}.
All our measurements of $\SWW$ 
are compared with the SM expectation in Figure~\ref{fig:xsec}.  
Good agreement is observed.

%
\section*{Acknowledgements}
We wish to 
express our gratitude to the CERN accelerator divisions for
the excellent performance of the LEP machine. 
We acknowledge the contributions of the engineers 
and technicians who have participated in the construction 
and maintenance of this experiment.  

\clearpage

%
\bibliographystyle{l3stylem}

\begin{mcbibliography}{10}

\bibitem{standard_model}
S.~L. Glashow, \NP {\bf 22} (1961) 579;\\ S. Weinberg, \PRL {\bf 19} (1967)
  1264;\\ A. Salam, in {\em Elementary Particle Theory}, ed. N. Svartholm,
  Stockholm, Alm\-quist and Wiksell (1968), 367\relax
\relax
\bibitem{SM-2}
M.~Veltman, \NP {\bf B7} (1968) 637;\\ G.M.~'t~Hooft, \NP {\bf B35} (1971)
  167;\\ G.M.~'t~Hooft and M.~Veltman, \NP {\bf B44} (1972) 189; \NP {\bf B50}
  (1972) 318\relax
\relax
\bibitem{CCNC}
D. Bardin \etal, Nucl. Phys. (Proc. Suppl.) {\bf B 37} (1994) 148;\\ F.A.
  Berends \etal, Nucl. Phys. (Proc. Suppl.) {\bf B 37} (1994) 163\relax
\relax
\bibitem{LEP2YRWW}
W. Beenakker \etal, in {\em Physics at LEP 2}, Report CERN 96-01 (1996), eds G.
  Altarelli, T. Sj{\"o}strand, F. Zwirner, Vol. 1, p. 79\relax
\relax
\bibitem{LEP2YREG}
D. Bardin \etal, in {\em Physics at LEP 2}, Report CERN 96-01 (1996), eds G.
  Altarelli, T. Sj{\"o}strand, F. Zwirner, Vol. 2, p. 3\relax
\relax
\bibitem{l3-01-new}
The L3 Collaboration, B. Adeva \etal, Nucl. Instr. and Meth. {\bf A 289} (1990)
  35; \\ M. Chemarin \etal, Nucl. Instr. and Meth. {\bf A 349} (1994) 345; \\
  M. Acciarri \etal, Nucl. Instr. and Meth. {\bf A 351} (1994) 300; \\ G. Basti
  \etal, Nucl. Instr. and Meth. {\bf A 374} (1996) 293; \\ A. Adam \etal, Nucl.
  Instr. and Meth. {\bf A 383} (1996) 342\relax
\relax
\bibitem{l3-lumi96}
I.C. Brock \etal, Nucl. Instr. and Meth. {\bf A 381} (1996) 236\relax
\relax
\bibitem{l3-196}
The L3 Collaboration, M.\ Acciarri \etal,
\newblock  Phys. Lett. {\bf B 479}  (2000) 101\relax
\relax
\bibitem{L-XSEC-161-183}
The L3 Collaboration, M. Acciarri \etal, Phys. Lett. {\bf B 398} (1997) 223;
  Phys. Lett. {\bf B 407} (1997) 419; Phys. Lett. {\bf B 436} (1998) 437\relax
\relax
\bibitem{GEOMJETS}
H.J. Daum \etal, Z. Phys. {\bf C 8} (1981) 167\relax
\relax
\bibitem{DURHAM}
S. Catani \etal, Phys. Lett. {\bf B 269} (1991) 432;\\ S. Bethke \etal, Nucl.
  Phys. {\bf B 370} (1992) 310\relax
\relax
\bibitem{KORALW}
KORALW version 1.33 is used.\\ S. Jadach \etal, Comp. Phys. Comm. {\bf 94}
  (1996) 216;\\ S. Jadach \etal, Phys. Lett. {\bf B 372} (1996) 289\relax
\relax
\bibitem{HERWIG}
HERWIG version 5.9 is used.\\ G.~Marchesini and B.~Webber, Nucl. Phys. {\bf B
  310} (1988) 461;\\ I.G. Knowles, Nucl. Phys. {\bf B 310} (1988) 571; \\ G.
  Marchesini $\etal$, Comp. Phys. Comm. {\bf 67} (1992) 465\relax
\relax
\bibitem{ARIADNE}
L.~L{\"o}nnblad,
\newblock  Comp. Phys. Comm. {\bf 71}  (1992) 15\relax
\relax
\bibitem{EXCALIBUR}
F.A. Berends, R. Kleiss and R. Pittau, \CPC {\bf 85} (1995) 437\relax
\relax
\bibitem{PYTHIA}
PYTHIA versions 5.722 and 6.1 are used.\\ T. Sj{\"o}strand, {\em PYTHIA~5.7 and
  JETSET~7.4 Physics and Manual}, \\ CERN-TH/7112/93 (1993), revised August
  1995; \CPC {\bf 82} (1994) 74; hep-ph/0001032\relax
\relax
\bibitem{KK2F}
S. Jadach, B.F.L. Ward and Z. W\c{a}s,
\newblock  Phys. Lett. {\bf B 449}  (1999) 97\relax
\relax
\bibitem{KORALZ}
KORALZ version 4.02 is used. \\ S. Jadach, B.F.L. Ward and Z. W\c{a}s, \CPC
  {\bf 79} (1994) 503\relax
\relax
\bibitem{BHAGENE}
J.H.~Field, \PL {\bf B 323} (1994) 432; \\ J.H.~Field and T.~Riemann, \CPC {\bf
  94} (1996) 53\relax
\relax
\bibitem{BHWIDE}
BHWIDE version 1.01 is used.\\ S.~Jadach, W.~Placzek, B.F.L.~Ward, Phys. Rev.
  {\bf D 40} (1989) 3582, Comp. Phys. Comm. {\bf 70} (1992) 305, \PL {\bf B
  390} (1997) 298\relax
\relax
\bibitem{TEEGG}
D.~Karlen, {\NP} {\bf B 289} (1987) 23\relax
\relax
\bibitem{DIAG36}
F.~A.~Berends, P.~H.~Daverfeldt and R. Kleiss,
\newblock  Nucl. Phys. {\bf B 253}  (1985) 441\relax
\relax
\bibitem{PHOJET}
PHOJET version 1.05 is used. \\ R.~Engel, \ZfP {\bf C 66} (1995) 203; R.~Engel
  and J.~Ranft, \PR {\bf D 54} (1996) 4244\relax
\relax
\bibitem{xsigel}
The L3 detector simulation is based on GEANT Version 3.15.\\ R. Brun \etal,
  {\em GEANT 3}, CERN-DD/EE/84-1 (Revised), 1987.\\ The GHEISHA program (H.
  Fesefeldt, RWTH Aachen Report PITHA 85/02 (1985)) \\ is used to simulate
  hadronic interactions\relax
\relax
\bibitem{LEP1YRSP}
Z. Kunszt \etal, in {\em Z Physics at LEP 1}, Report CERN 89-08 (1989), eds
  G.~Altarelli, R.~Kleiss, C.~Verzegnassi, Vol. 1, p. 385\relax
\relax
\bibitem{PDG98}
C. Caso \etal, Eur. Phys. J. {\bf C 3} (1998) 1\relax
\relax
\bibitem{ISR}
G.J. van Oldenborgh, \NP {\bf B 470} (1996) 71\relax
\relax
\bibitem{PHOTOS}
E.~Barberio and Z. W\c{a}s, \CPC {\bf 79} (1994) 291;\\ E.~Barberio, B. van
  Eijk and Z. W\c{a}s, \CPC {\bf 66} (1991) 115\relax
\relax
\bibitem{l3-213}
The L3 Collaboration, M.\ Acciarri \etal, {``}Measurement of Bose-Einstein
  Correlations in $\mathrm{e^+e^-}$ $\rightarrow $$\mathrm{W^+ W^-}$ at
  $\sqrt{s} \approx 189$ GeV{''}, submitted to Phys. Lett. {\bf B}\relax
\relax
\bibitem{LUBOEI}
L.~L{\"o}nnblad and T.~Sj{\"o}strand, Eur. Phys. J. {\bf C 2} (1998) 165\relax
\relax
\bibitem{SKMODELS}
T. Sj{\"o}strand and V.~Khoze, Z. Phys. {\bf C 64} (1994) 281; Eur. Phys. J.
  {\bf C 6} (1999) 271\relax
\relax
\bibitem{GENTLE}
GENTLE version 2.0 is used.\\ D. Bardin \etal, Comp. Phys. Comm. {\bf 104}
  (1997) 161\relax
\relax
\bibitem{RACONWW}
A.~Denner \etal, \PL {\bf B 475} (2000) 127; hep-ph/0006307\relax
\relax
\bibitem{YFSWW3}
YFSWW3 version 1.14 is used. \\ S.~Jadach \etal, \PR {\bf D 54} (1996) 5434;
  Phys. Lett. {\bf B 417} (1998) 326; \PR {\bf D 61} (2000) 113010;
  hep-ph/0007012\relax
\relax
\bibitem{DPA}
W.~Beenakker, F.A.~Berends and A.P.~Chapovsky,
\newblock  Nucl. Phys. {\bf B 548}  (1999) 3\relax
\relax
\bibitem{LEP2MC4F}
M.W. Gr{\"u}newald \etal, hep-ph/0005309\relax
\relax
\bibitem{VCKM}
N. Cabibbo, Phys. Rev. Lett. {\bf 10} (1963) 531;\\ M. Kobayashi and T.
  Maskawa, Prog. Theor. Phys. {\bf 49} (1973) 652\relax
\relax
\bibitem{ADXSEC-189}
The DELPHI Collaboration, P. Abreu \etal, CERN-EP/2000-035; \\ The ALEPH
  Collaboration, R. Barate \etal, CERN-EP/2000-052\relax
\relax
\end{mcbibliography}

%
%
\newpage
\typeout{   }     
\typeout{Using author list for paper 218 ONLY }
\typeout{Using author list for paper 218 ONLY }
\typeout{Using author list for paper 218 ONLY }
\typeout{Using author list for paper 218 ONLY }
\typeout{Using author list for paper 218 ONLY }
\typeout{Using author list for paper 218 ONLY }
\typeout{$Modified: Fri Jul 21 15:51:14 2000 by clare $}
\typeout{!!!!  This should only be used with document option a4p!!!!}
\typeout{   }
%
%
%
%
%
%

\newcount\tutecount  \tutecount=0
\def\tutenum#1{\global\advance\tutecount by 1 \xdef#1{\the\tutecount}}
\def\tute#1{$^{#1}$}
\tutenum\aachen            
\tutenum\nikhef            
\tutenum\mich              
\tutenum\lapp              
\tutenum\basel             
\tutenum\lsu               
\tutenum\beijing           
\tutenum\berlin            
\tutenum\bologna           
\tutenum\tata              
\tutenum\ne                
\tutenum\bucharest         
\tutenum\budapest          
\tutenum\mit               
\tutenum\debrecen          
\tutenum\florence          
\tutenum\cern              
\tutenum\wl                
\tutenum\geneva            
\tutenum\hefei             
\tutenum\seft              
\tutenum\lausanne          
\tutenum\lecce             
\tutenum\lyon              
\tutenum\madrid            
\tutenum\milan             
\tutenum\moscow            
\tutenum\naples            
\tutenum\cyprus            
\tutenum\nymegen           
\tutenum\caltech           
\tutenum\perugia           
\tutenum\cmu               
\tutenum\prince            
\tutenum\rome              
\tutenum\peters            
\tutenum\potenza           
\tutenum\salerno           
\tutenum\ucsd              
\tutenum\santiago          
\tutenum\sofia             
\tutenum\korea             
\tutenum\alabama           
\tutenum\utrecht           
\tutenum\purdue            
\tutenum\psinst            
\tutenum\zeuthen           
\tutenum\eth               
\tutenum\hamburg           
\tutenum\taiwan            
\tutenum\tsinghua          

{
\parskip=0pt
\noindent
{\bf The L3 Collaboration:}
\ifx\selectfont\undefined
 \baselineskip=10.8pt
 \baselineskip\baselinestretch\baselineskip
 \normalbaselineskip\baselineskip
 \ixpt
\else
 \fontsize{9}{10.8pt}\selectfont
\fi
\medskip
\tolerance=10000
\hbadness=5000
\raggedright
\hsize=162truemm\hoffset=0mm
\def\r{\rlap,}
\noindent

M.Acciarri\r\tute\milan\
P.Achard\r\tute\geneva\ 
O.Adriani\r\tute{\florence}\ 
M.Aguilar-Benitez\r\tute\madrid\ 
J.Alcaraz\r\tute\madrid\ 
G.Alemanni\r\tute\lausanne\
J.Allaby\r\tute\cern\
A.Aloisio\r\tute\naples\ 
M.G.Alviggi\r\tute\naples\
G.Ambrosi\r\tute\geneva\
% ...
\newpage
\begin{list}{A}{\itemsep=0pt plus 0pt minus 0pt\parsep=0pt plus 0pt minus 0pt
                \topsep=0pt plus 0pt minus 0pt}
\item[\aachen]
 I. Physikalisches Institut, RWTH, D-52056 Aachen, FRG$^{\S}$\\
 III. Physikalisches Institut, RWTH, D-52056 Aachen, FRG$^{\S}$
\item[\nikhef] National Institute for High Energy Physics, NIKHEF, 
     and University of Amsterdam, NL-1009 DB Amsterdam, The Netherlands
\item[\mich] University of Michigan, Ann Arbor, MI 48109, USA
\item[\lapp] Laboratoire d'Annecy-le-Vieux de Physique des Particules, 
     LAPP,IN2P3-CNRS, BP 110, F-74941 Annecy-le-Vieux CEDEX, France
\item[\basel] Institute of Physics, University of Basel, CH-4056 Basel,
     Switzerland
\item[\lsu] Louisiana State University, Baton Rouge, LA 70803, USA
\item[\beijing] Institute of High Energy Physics, IHEP, 
  100039 Beijing, China$^{\triangle}$ 
\item[\berlin] Humboldt University, D-10099 Berlin, FRG$^{\S}$
\item[\bologna] University of Bologna and INFN-Sezione di Bologna, 
     I-40126 Bologna, Italy
\item[\tata] Tata Institute of Fundamental Research, Bombay 400 005, India
\item[\ne] Northeastern University, Boston, MA 02115, USA
\item[\bucharest] Institute of Atomic Physics and University of Bucharest,
     R-76900 Bucharest, Romania
\item[\budapest] Central Research Institute for Physics of the 
     Hungarian Academy of Sciences, H-1525 Budapest 114, Hungary$^{\ddag}$
\item[\mit] Massachusetts Institute of Technology, Cambridge, MA 02139, USA
\item[\debrecen] KLTE-ATOMKI, H-4010 Debrecen, Hungary$^\P$
\item[\florence] INFN Sezione di Firenze and University of Florence, 
     I-50125 Florence, Italy
\item[\cern] European Laboratory for Particle Physics, CERN, 
     CH-1211 Geneva 23, Switzerland
\item[\wl] World Laboratory, FBLJA  Project, CH-1211 Geneva 23, Switzerland
\item[\geneva] University of Geneva, CH-1211 Geneva 4, Switzerland
\item[\hefei] Chinese University of Science and Technology, USTC,
      Hefei, Anhui 230 029, China$^{\triangle}$
\item[\seft] SEFT, Research Institute for High Energy Physics, P.O. Box 9,
      SF-00014 Helsinki, Finland
\item[\lausanne] University of Lausanne, CH-1015 Lausanne, Switzerland
\item[\lecce] INFN-Sezione di Lecce and Universit\'a Degli Studi di Lecce,
     I-73100 Lecce, Italy
\item[\lyon] Institut de Physique Nucl\'eaire de Lyon, 
     IN2P3-CNRS,Universit\'e Claude Bernard, 
     F-69622 Villeurbanne, France
\item[\madrid] Centro de Investigaciones Energ{\'e}ticas, 
     Medioambientales y Tecnolog{\'\i}cas, CIEMAT, E-28040 Madrid,
     Spain${\flat}$ 
\item[\milan] INFN-Sezione di Milano, I-20133 Milan, Italy
\item[\moscow] Institute of Theoretical and Experimental Physics, ITEP, 
     Moscow, Russia
\item[\naples] INFN-Sezione di Napoli and University of Naples, 
     I-80125 Naples, Italy
\item[\cyprus] Department of Natural Sciences, University of Cyprus,
     Nicosia, Cyprus
\item[\nymegen] University of Nijmegen and NIKHEF, 
     NL-6525 ED Nijmegen, The Netherlands
\item[\caltech] California Institute of Technology, Pasadena, CA 91125, USA
\item[\perugia] INFN-Sezione di Perugia and Universit\'a Degli 
     Studi di Perugia, I-06100 Perugia, Italy   
\item[\cmu] Carnegie Mellon University, Pittsburgh, PA 15213, USA
\item[\prince] Princeton University, Princeton, NJ 08544, USA
\item[\rome] INFN-Sezione di Roma and University of Rome, ``La Sapienza",
     I-00185 Rome, Italy
\item[\peters] Nuclear Physics Institute, St. Petersburg, Russia
\item[\potenza] INFN-Sezione di Napoli and University of Potenza, 
     I-85100 Potenza, Italy
\item[\salerno] University and INFN, Salerno, I-84100 Salerno, Italy
\item[\ucsd] University of California, San Diego, CA 92093, USA
\item[\santiago] Dept. de Fisica de Particulas Elementales, Univ. de Santiago,
     E-15706 Santiago de Compostela, Spain
\item[\sofia] Bulgarian Academy of Sciences, Central Lab.~of 
     Mechatronics and Instrumentation, BU-1113 Sofia, Bulgaria
\item[\korea]  Laboratory of High Energy Physics, 
     Kyungpook National University, 702-701 Taegu, Republic of Korea
\item[\alabama] University of Alabama, Tuscaloosa, AL 35486, USA
\item[\utrecht] Utrecht University and NIKHEF, NL-3584 CB Utrecht, 
     The Netherlands
\item[\purdue] Purdue University, West Lafayette, IN 47907, USA
\item[\psinst] Paul Scherrer Institut, PSI, CH-5232 Villigen, Switzerland
\item[\zeuthen] DESY, D-15738 Zeuthen, 
     FRG
\item[\eth] Eidgen\"ossische Technische Hochschule, ETH Z\"urich,
     CH-8093 Z\"urich, Switzerland
\item[\hamburg] University of Hamburg, D-22761 Hamburg, FRG
\item[\taiwan] National Central University, Chung-Li, Taiwan, China
\item[\tsinghua] Department of Physics, National Tsing Hua University,
      Taiwan, China
\item[\S]  Supported by the German Bundesministerium 
        f\"ur Bildung, Wissenschaft, Forschung und Technologie
\item[\ddag] Supported by the Hungarian OTKA fund under contract
numbers T019181, F023259 and T024011.
\item[\P] Also supported by the Hungarian OTKA fund under contract
  numbers T22238 and T026178.
\item[$\flat$] Supported also by the Comisi\'on Interministerial de Ciencia y 
        Tecnolog{\'\i}a.
\item[$\sharp$] Also supported by CONICET and Universidad Nacional de La Plata,
        CC 67, 1900 La Plata, Argentina.
\item[$\diamondsuit$] Also supported by Panjab University, Chandigarh-160014, 
        India.
\item[$\triangle$] Supported by the National Natural Science
  Foundation of China.
\end{list}
}
\vfill


\newpage

\begin{table}[p]
\begin{center}
\renewcommand{\arraystretch}{1.5}
\begin{tabular}{| l || c c c c c c | c c c | c|}
\hline
 & 
\multicolumn{10}{|c|}{Efficiencies [\%] for} \\
\multicolumn{1}{|c||}{Selection}      
          &  $\ENEN$ &  $\ENMN$ &  $\ENTN$ &  $\MNMN$ &  $\MNTN$ & 
             $\TNTN$ &  $\QQEN$ &  $\QQMN$ &  $\QQTN$ &  $\QQQQ$ \\
\hline
\hline
$\EEENENG$&  $64.2$  & $\pz0.3$ & $12.6 $  &  ---     &  ---     & 
            $\pz1.1$ &          &          &          &          \\
$\EEENMNG$&   ---    &  $60.6$  & $10.0$   & $\pz0.5$ & $10.8$   & 
            $\pz1.7$ &          &          &          &          \\
$\EEENTNG$& $\pz5.3$ & $\pz1.1$ & $36.8$   &  ---     & $\pz0.1$ & 
            $\pz7.1$ &          &          &          &          \\
$\EEMNMNG$&   ---    & ---      &  ---     &  $50.6$  & $\pz8.7$ & 
            $\pz0.8$ &          &          &          &          \\
$\EEMNTNG$&   ---    & $\pz2.9$ & $\pz0.2$ & $\pz2.2$ &  $31.6$  & 
            $\pz4.7$ &          &          &          &          \\
$\EETNTNG$& $\pz0.1$ & $\pz0.1$ & $\pz1.3$ &   ---    & $\pz0.8$ & 
             $19.8$  &          &          &          &          \\
\hline
$\EEQQENG$&          &          &          &          &          & 
                     &  $81.5$  & $\pz0.3$ & $\pz1.7$ &  ---     \\
$\EEQQMNG$&          &          &          &          &          & 
                     & $\pz0.2$ &  $76.7$  & $\pz3.6$ &  ---     \\
$\EEQQTNG$&          &          &          &          &          & 
                     & $\pz5.3$ & $\pz6.5$ &  $50.6$  & $\pz0.3$ \\
\hline
$\EEQQQQG$&          &          &          &          &          & 
                     & $\pz0.2$ &  ---     & $\pz0.8$ &  $87.2$  \\
\hline
\end{tabular}
\caption[]{
  Selection efficiencies for signal processes $\EELNLNG$, $\EEQQLNG$,
  and $\EEQQQQG$.  For the $\LNLN$ and $\QQEN$ selections, the signal
  efficiencies are  derived from a CC56+NC56 and a CC20 Monte Carlo sample,
  respectively,  and
  given within phase-space cuts.  
  The total efficiencies at CC03 level for the
  $\LNLN$ and $\QQEN$ selections are 52.3\% and 78.1\%, respectively.
  For the $\QQQQ$ selection, the numbers are quoted for a
  neural-network output larger than 0.6.}
\label{tab:xmat-1}
\end{center}
\end{table}

\begin{table}[p]
\begin{center}
\renewcommand{\arraystretch}{1.5}
\begin{tabular}{| l || c | c |}
\hline
\multicolumn{1}{|c||}{Selection} & 
{$N_{\mathrm{data}}$} & {$N_{\mathrm{bg}}$} \\
\hline
\hline
$\EEENENG$& $\pzz49$ & $\pzz9.4\pm 3.0$  \\
$\EEENMNG$& $\pzz43$ & $\pzz4.2\pm 0.7$  \\
$\EEENTNG$& $\pzz38$ & $\pzz2.3\pm 0.7$  \\
$\EEMNMNG$& $\pzz24$ & $\pzz8.3\pm 0.9$  \\
$\EEMNTNG$& $\pzz26$ & $\pzz0.9\pm 0.2$  \\
$\EETNTNG$& $\pzz10$ & $\pzz1.1\pm 0.4$  \\
\hline                                  
$\EEQQENG$& $\pz363$ & $\pz14.8\pm 0.9$  \\
$\EEQQMNG$& $\pz340$ & $\pz13.8\pm 0.7$  \\
$\EEQQTNG$& $\pz329$ & $\pz41.7\pm 0.7$  \\
\hline                                  
$\EEQQQQG$& $  1431$ & $266.6\pm 1.2$    \\
\hline
\end{tabular}
\caption[]{
  Number of selected data events, $N_{\mathrm{data}}$, and  number of
  expected non-WW background events, $N_{\mathrm{bg}}$, for the
  different selections.
  The uncertainties include only the Monte Carlo statistics. 
  For the $\QQQQ$ selection, the numbers are quoted for a
  neural-network output larger than 0.6.
}
\label{tab:xmat-2}
\end{center}
\end{table}

\begin{table}[p]
\begin{center}
\renewcommand{\arraystretch}{1.15}
\begin{tabular}{|l||r|r|r|r|r|}
\hline
\multicolumn{6}{|c|}{Systematic Uncertainties on $\sigma~[\%]$}\\
\hline
& \multicolumn{5}{|c|}{Final State} \\
\cline{2-6}
Source                      &$\LNLN$&$\QQEN$&$\QQMN$&$\QQTN$&$\QQQQ$\\
\hline
Luminosity                  & \multicolumn{5}{c|}{0.2}        \\
\cline{2-6}
MC statistics (signal)      &   0.6 &  0.2 &  0.3 &  0.5 &  0.5 \\
MC statistics (background)  &   2.0 &  0.1 &  0.2 &  0.4 &  0.1 \\
Selection procedure         &   1.4 &  1.2 &  1.3 &  2.5 &  1.1 \\
Background cross sections   &  \multicolumn{1}{l|}{$< 0.1$} &  0.1 &  0.3 &  0.3 &  0.2 \\
W mass  ($\pm0.10~\GeV$)    &   0.3 &  0.1 &  0.1 &  0.2 &  0.1 \\ 
W width ($\pm0.06~\GeV$)    &   0.2 &  0.1 & \multicolumn{1}{l|}{$< 0.1$}&  0.1 & \multicolumn{1}{l|}{$< 0.1$} \\
\cline{3-6}
ISR simulation              &   0.2 & \multicolumn{4}{c|}{$<0.1$} \\ 
\cline{3-6}
FSR simulation              &   0.4 &  0.9 &  0.7 &  0.3 &  0.2 \\ 
CC03 versus 4F              &   --- &  --- &  0.1 &  0.1 &  0.4 \\ 
Hadronisation (signal)   &   --- &  0.3 &  0.3 &  1.1 &  1.8 \\ 
Hadronisation (background) &   --- &  --- &  --- &  --- &  0.6 \\ 
\cline{3-5}
Bose-Einstein effects       &   --- &  \multicolumn{3}{c|}{$<0.1$}   &  0.3 \\ 
\cline{3-5}
Colour Reconnection         &   --- &  --- &  --- &  --- &  0.3 \\ 
\hline
Total                       &   2.6 &  1.5 &  1.6 &  2.9 &  2.3 \\
\hline
\end{tabular}
\caption[]{
  Contributions to the systematic uncertainty on the cross section
  measurements.  The systematic uncertainties are relative to the
  cross sections listed in Table~\ref{tab:xsec}. 
\label{tab:syst}
}
\end{center}
\end{table}

\begin{table}[p]
\begin{center}
\renewcommand{\arraystretch}{1.5}
\begin{tabular}{| l || c | c || c |}
\hline
          & {Conversion}         
          & $\sigma(\mathrm{CC03})$
          & $\sigma_{\mathrm{SM}} $ 
          \\
\multicolumn{1}{|c||}{Process}  & {factor $f$} & [pb] & [pb] \\
\hline\hline
$\EEENENG$&  $0.88$& $0.22\pm0.06\pm0.01$& 0.19 \\      
$\EEENMNG$&  $1.07$& $0.22\pm0.07\pm0.01$& 0.37 \\   
$\EEENTNG$&  $1.07$& $0.50\pm0.11\pm0.01$& 0.37 \\   
$\EEMNMNG$&  $0.96$& $0.10\pm0.06\pm0.01$& 0.19 \\   
$\EEMNTNG$&  $1.10$& $0.43\pm0.11\pm0.01$& 0.37 \\   
$\EETNTNG$&  $0.96$& $0.20\pm0.09\pm0.01$& 0.19 \\   
\hline                                          
$\EEQQENG$&  $1.01$& $2.39\pm0.13\pm0.04$& 2.37 \\   
$\EEQQMNG$&    --- & $2.27\pm0.14\pm0.04$& 2.37 \\   
$\EEQQTNG$&    --- & $2.64\pm0.21\pm0.08$& 2.37 \\   
\hline                                          
$\EEQQQQG$&    --- & $7.36\pm0.24\pm0.18$& 7.41 \\   
\hline
\end{tabular}
\caption[]{
  Conversion factors and cross sections of four-fermion final states.
  The ratio $f$ of the CC03 cross section without cuts and the
  four-fermion cross section within phase-space cuts is calculated
  with EXCALIBUR and listed in the second column.  The third column
  shows the measured CC03 cross sections, $\sigma(\mathrm{CC03})$.
  The $\QQQQ$ cross section is obtained from a fit to the
  neural-network output distribution as described in
  Section~\ref{sec:qqqq}.  The first uncertainty is statistical and the
  second systematic.  Also shown are the SM predictions for the 
  cross sections, $\sigma_{\mathrm{SM}}$.

\label{tab:xsec}
}
\end{center}
\end{table}

\begin{table}[p]
\begin{center}
\renewcommand{\arraystretch}{1.5}
\begin{tabular}{| l || c c c c c c | c c c c|}
\hline
\multicolumn{11}{|c|}{Correlation coefficients} \\
\hline
          &  $\ENEN$ &  $\ENMN$ &  $\ENTN$ &  $\MNMN$ &  $\MNTN$ & 
             $\TNTN$ &  $\QQEN$ &  $\QQMN$ &  $\QQTN$ &  $\QQQQ$ \\
\hline
\hline
$\ENEN$&$+1.00$&$+0.09$&$-0.37$&$\pz0.00$&$-0.01$&$+0.05$&&&& \\
$\ENMN$&       &$+1.00$&$-0.27$&  $+0.09$&$-0.31$&$+0.07$&&&& \\
$\ENTN$&       &       &$+1.00$&  $-0.01$&$+0.05$&$-0.25$&&&& \\
$\MNMN$&       &       &       &  $+1.00$&$-0.32$&$+0.03$&&&& \\
$\MNTN$&       &       &       &         &$+1.00$&$-0.18$&&&& \\
$\TNTN$&       &       &       &         &       &$+1.00$&&&& \\
\hline
$\QQEN$&&&&&&&  $+1.00$&$\pz0.00$&$-0.10$ &$\pz0.00$\\
$\QQMN$&&&&&&&         &  $+1.00$&$-0.15$ &$\pz0.00$\\
$\QQTN$&&&&&&&         &         &$+1.00$ &$-0.01$  \\
$\QQQQ$&&&&&&&         &         &        &$+1.00$  \\
\hline
\end{tabular}

\caption[]{Correlation coefficients between the cross section
  measurements of single channels.
\label{tab:corrmatrix}
}
\end{center}
\end{table}

\begin{table}[p]
\begin{center}
\renewcommand{\arraystretch}{1.5}
\begin{tabular}{| c || c | c || c |}
\hline
          &          Lepton          &    Lepton            & Standard \\
Parameter &        Non-Universality  &  Universality        & Model    \\
\hline\hline
$B(\WEN)~[\%]$& $   10.77\pm0.45\pm0.16$ &      ---             &  ---    \\
$B(\WMN)~[\%]$& $\pz 9.90\pm0.46\pm0.15$ &      ---             &  ---    \\
$B(\WTN)~[\%]$& $   11.24\pm0.62\pm0.22$ &      ---             &  ---    \\
$B(\WLN)~[\%]$&        ---               & $10.60\pm0.23\pm 0.11$   &
$10.83$ \\
$B(\WQQ)~[\%]$& $   68.09\pm0.69\pm0.33$ & $68.20\pm0.68\pm 0.33$   &
$67.51$ \\
\hline
$\SWW~[\mathrm{pb}]$ 
              & $16.29\pm0.37\pm0.22$ & $16.22\pm0.37\pm 0.21$ & $16.24$ \\
\hline
\end{tabular}
\vskip 0.5cm
\caption[]{W-decay branching fractions, $B$, and total W-pair cross 
  section, $\mathrm{\sigma_{WW}}$, derived without and with the
  assumption of charged-current lepton universality.  The correlations
  between the leptonic branching fractions are $-0.03$, $-0.28$ and
  $-0.29$ for e$\mu$, e$\tau$ and  $\mu\tau$, respectively.  
  Also shown are the W-decay branching fractions~\cite{LEP2YRWW} and
  the total W-pair cross section as expected in the SM.
\label{tab:brafra}
}
\end{center}
\end{table}

\clearpage

\clearpage

\clearpage

\clearpage

\clearpage

\end{document}